\documentstyle[11pt,epsf]{article}
\newcommand{\ewxy}[2]{\setlength{\epsfxsize}{#2}\epsfbox[10 60 640 570]{#1}}

\def\to{\rightarrow}

\let\<=\langle
\let\>=\rangle

\def\eqn#1{(\ref{#1})}  %Sticks parentheses around the LaTex "ref" macro
\def\beq{\begin{equation}}
\def\eeq{\end{equation}}
\def\ba{\begin{array}}
\def\bea{\begin{eqnarray}}
\def\ea{\end{array}}
\def\eea{\end{eqnarray}}

\def\comment#1{ \hbox{[{\it Comment suppressed here.}\/]} }
\def\hide#1{}
\def\o{\over}

\def\Tr{\hbox{Tr}}

\def\Ord{ {\rm O} }

\def\Re{ {\rm Re}\, }
\def\a{{\bf a}}

\def\r{{\bf r}}

\def\IR{\relax{\rm I\kern-.18em R}}
\def\IN{\relax{\rm I\kern-.18em N}}
\def\IB{\relax{\rm I\kern-.18em B}}
\def\IE{\relax{\rm I\kern-.18em E}}
\def\ZZ{\relax{\sf Z\kern-.4em Z}}

\def\TT{\mathchoice
       {\sf T\kern-0.52 em{T}}{\sf T\kern-0.52 em{T}}
       {\sf T\kern-0.40 em{T}}{\sf T\kern-0.40 em{T}}}
\def\IP{\mathchoice
       {\sf I\kern-0.14 em{P}}{\sf I\kern-0.14 em{P}}
       {\sf I\kern-0.11 em{P}}{\sf I\kern-0.11 em{P}}}
\def\id{1\kern-.25em {\rm l}}

\def\ontopss#1#2#3#4{\raise#4ex \hbox{#1}\mkern-#3mu {#2}}

\newcommand{\skipover}[1]{}

\def\s={\! = \!}        % for eqs in text

\def\l={\,=\,}
\def\+{\,+\,}
\def\-{\,-\,}

\newdimen\pmboffset
\def\oldpmb#1{\setbox0=\hbox{#1}%
 \copy0\kern-\wd0
 \kern\pmboffset\raise 1.732\pmboffset\copy0\kern-\wd0
 \kern\pmboffset\box0}

\def\appendix{\par                              % Have \appendix say
    \setcounter{section}{0}                     % `Appendix A', not just `A'
    \setcounter{subsection}{0}
    \renewcommand{\theequation}{\Alph{section}.\arabic{equation}}
    \renewcommand{\thesection}{Appendix \Alph{section}
                \setcounter{equation}{0}  } %Have eqns numbered (A.1) etc
}

\catcode`\@=11

\def\section{
\setcounter{equation}{0}        % Reset eqn numbers at start of section
\@startsection {section}{1}{\z@}{-3.5ex plus -1ex minus 
  -.2ex}{2.3ex plus .2ex}{\Large\bf}}
\renewcommand{\theequation}{\arabic{section}.\arabic{equation}}

\def\subsection{\@startsection{subsection}{2}{\z@}{-3.25ex plus -1ex minus 
 -.2ex}{1.5ex plus .2ex}{\normalsize\bf}}

\def\subsubsection{\@startsection{subsubsection}{3}{\z@}{-3.25ex plus
 -1ex minus -.2ex}{1.5ex plus .2ex}{\normalsize}}

\def\@eqnnum{%
\savebox{\eqnumb}{\rm (\theequation)}%
\settowidth{\numblen}{\usebox{\eqnumb}}%
\makebox[\numblen][l]{\usebox{\eqnumb}~~~\usebox{\eqlabel}}}

\catcode`\@=12

\newsavebox{\eqlabel}

\catcode`\@=11
\newlength{\numblen}
\newsavebox{\eqnumb}
\def\@eqnnum{%
\savebox{\eqnumb}{\rm (\theequation)}%
\settowidth{\numblen}{\usebox{\eqnumb}}%
\makebox[\numblen][l]{\usebox{\eqnumb}~~~\usebox{\eqlabel}}%
}
\catcode`\@=12

\newenvironment{equationwithlabel}[1]{ %
  \savebox{\eqlabel}{#1}
  \begin{equation}\label{#1} }{\end{equation}\savebox{\eqlabel}{~}}

\newcommand{\beql}[1]{\begin{equationwithlabel}{#1}}
\newcommand{\eeql}{\end{equationwithlabel}}

\newenvironment{eqnarraywithlabel}[1]{ %
  \savebox{\eqlabel}{#1}
  \begin{eqnarray}\label{#1} }{\end{eqnarray}\savebox{\eqlabel}{~}}

\newcommand{\beal}[1]{\begin{eqnarraywithlabel}{#1}}
\newcommand{\eeal}{\end{eqnarraywithlabel}}
\begin{document}

\begin{flushright}
FSU-SCRI-98-22 \\
March 1998 \\
% March 1998 (revised) \\
\end{flushright}
\vskip 6mm
\begin{center}

{\LARGE \bf The Anisotropic Wilson Gauge Action}

\vskip 9mm
{\normalsize Timothy R. Klassen\\[1.2mm]
             SCRI, Florida State University\\[1mm]
             Tallahassee, FL 32306-4130, USA}
\vskip 8mm

{\normalsize \bf Abstract}

\vskip5mm

\begin{minipage}{5.0in}  
{\small 
Anisotropic lattices, with a temporal lattice spacing smaller than
the spatial one, allow  precision Monte Carlo 
calculations of problems that are difficult to study otherwise:
heavy quarks, glueballs, hybrids, % large momentum form factors, % need small a_s
and high temperature thermodynamics, for example.  
We here perform the first step required for such 
studies with the (quenched) Wilson gauge action, namely, the determination 
of the renormalized anisotropy $\xi$ as a function of the bare anisotropy
$\xi_0$ and the coupling.  By, % roughly speaking, 
essentially,     comparing the finite-volume heavy quark potential
where the quarks are separated along a spatial direction with that where they 
are separated along the time direction, we determine the relation between
$\xi$ and $\xi_0$ to a fraction of 1\% for weak and to 1\% for strong
coupling.  We present a simple   parameterization of this relation for
% $1\leq \xi \leq 4$  and $5.5 \leq \beta \leq \infty$,  TK2:
$1\leq \xi \leq 6$  and $5.5 \leq \beta \leq \infty$, 
which incorporates the known one-loop result
and reproduces our non-perturbative determinations within errors.
% The ultimate aim of this project is the extension of the non-perturbative
% O($a$) improvement program to anisotropic lattices.
Besides solving the problem of how to choose the bare anisotropies if one 
wants to take the continuum limit at fixed renormalized anisotropy, this
parameterization also yields accurate estimates of the derivative
$\partial\xi_0/\partial\xi$ needed in thermodynamic studies.
}

\end{minipage}
\end{center}
\vskip 8mm

% \newpage
\renewcommand{\thepage}{\arabic{page}}
\setcounter{page}{1}

%%%%%%%%%%%%%%%%%%%%%%%%%%%%%%%%%%%%%%%%%%%%%%%%%%%%%%%%%%%%%%%%%%%%%%%%%%%
%% Body of paper

\section{Introduction}

Lattice QCD has experienced rapid progress in the last few years
(see e.g.~the proceedings of the latest lattice conference~\cite{LAT97}).
In particular, it should be viewed as progress that we can now
clearly see the limitations of various approximations that have 
been commonly used in the past, like quenching or the use of
(mean-field improved) perturbative coefficients for the 
improvement terms in actions and currents.

A great theoretical advance in the last two years was the determination
of the {\it non-perturbatively} $\Ord(a)$ improved Wilson quark
action for both quenched~\cite{ALPHA,EHKlat97,EHKprl} and full ($n_f\s=2$)
QCD~\cite{JanSom} initiated by the ALPHA collaboration. These results
are most relevant for the study of light hadrons.
They are therefore only of limited value in the study of heavy quark
systems, where we are also beginning to see the limitations of
current methods. The most popular of these are the NRQCD~\cite{NRQCD} and 
Fermilab~\cite{EKM} approaches. The problem is, as for the standard  Wilson
quark action, that the non-perturbative corrections to the O($a$) improvement
coefficients in these actions are large on coarse lattices. Different 
mean-field prescriptions to take some of these corrections into account
give quite different results on coarse lattices~\cite{Trottier}.
A non-perturbative determination of these coefficients is therefore
called for.

We think that the most reliable and ultimately simplest approach is to
perform the non-perturbative O($a$) improvement {\it not} for the NRQCD or
Fermilab formalisms, but rather to extend the Symanzik improvement program
for Wilson-type quark actions to {\it anisotropic} lattices, with a 
temporal lattice spacing $a_t$ much smaller than the spatial one $a_s$.
On such lattices heavy quarks will not suffer large lattice artifacts
as long as their masses are small in units of $a_t$. 
At the expense of the relatively modest cost incured by simulating a lattice 
with a larger temporal extent one can study heavy quark systems in a 
relativistic framework with controlled errors.
Such studies would be orders of magnitude more expensive on isotropic lattices.
The use of anisotropic lattices for heavy quark physics 
was advocated in refs.~\cite{ILQA,AKLlat96,SF}; we refer to~\cite{ILQA}
for studies of the classical case and to~\cite{AKLlat96} for some exploratory 
simulations.

Anisotropic lattices are also important  in other situations. Generally
speaking, they allow one to reap many of the benefits of fine lattices,
while still using cheap coarse spatial lattices:

\begin{itemize}

\item
The signal to noise ratio of correlation functions calculated in Monte Carlo
simulations generically decays exponentially in time. Choosing a smaller
temporal lattice spacing gives more time slices with an accurate signal,
enabling more precise and confident mass determinations, for example.
This is crucial for particles with bad signal/noise properties like
glueballs~\cite{MorPea}, hybrids and P-state mesons.
For the same reason it is also clear that {\it excited-state}
masses can be determined much more accurately on anisotropic lattices.

\item
For a full determination of thermodynamic quantities one has to
be able to take independent derivatives with respect to temperature and
volume. The simplest way to achieve this, is to have independent spatial
and temporal lattice spacings. 
 ~The point of the previous paragraph
also becomes important at high temperatures, where one needs
a small temporal lattice spacing to have a sufficient number of Matsubara
frequencies for accurate studies of the electro-weak phase transition,
transport coefficients in the quark-gluon plasma phase, and many other issues.

\item
By euclidean invariance we can also think of the small lattice spacing
as a spatial one. 
% This opens the door to studies of systems where large
% spatial momenta are important, e.g.~large momentum form factors.
This allows one to study large spatial momenta, which are 
phenomenologically important for form factors, for example.

\end{itemize}

Of course, the advantages of anisotropic lattices come at a price,
namely, that coefficients in the action have to be
tuned to restore space-time exchange symmetry on the quantum level.
In the context of the O($a$) improvement program, the good news is that
the increase in the number of coefficients that have to be determined for
a relativistic action is quite modest:
For a Wilson-type quark action there are three (instead of one) coefficients 
to be tuned to achieve non-perturbative O($a$) improvement~\cite{SF}. 
How to achieve this using the Schr\"odinger functional, similar 
to the isotropic case~\cite{ALPHA}, has been outlined in~\cite{SF}.

However, before worrying about the quark action, we have to consider 
gauge actions on anisotropic lattices. The first step is the study of
pure gauge theory,  relevant for quenched QCD.  
In this paper we will consider the simplest case, the anisotropic Wilson gauge 
action~\cite{Karsch,Burgers}.\footnote{Determinations of the renormalized 
anisotropy for improved actions will be reported in~\cite{anisoQCD}.}
For this action no coefficients have to be tuned to
restore space-time exchange symmetry up to O($a^2$) errors.
Nevertheless, there is something to be done, since we have to
know the true or {\it renormalized anisotropy} $\xi\equiv a_s/a_t\,$ as a 
function of the bare parameters.  This is important, because 
we would like to take the continuum limit at a fixed value of 
the renormalized anisotropy, for example.

Our aim in this paper is to develop a simple and accurate method to determine
the renormalized anisotropy and apply it to the Wilson plaquette action. 
For SU($N$) gauge fields we write this action on an anisotropic lattice as
\beq\label{anisoW}
  S  ~=~ {\beta \o N} \, 
                 \sum_{x,s>s'} \, {1\o \xi_0} \, \Re\Tr\, (1-P_{ss'}(x)) \+
                                        \xi_0  \, \Re\Tr\, (1-P_{0s}(x)) \, .
\eeq
Here $P_{\mu\nu}(x)$ is the plaquette operator in the 
$\mu\nu$-plane,\footnote{Here a summary of our lattice conventions:
We always work on four-dimensional hypercubic lattices with periodic boundary 
conditions. Directions are labelled by $\mu=0,1,2,3$ or $t,s,s'$, where $t\s= 0$
stands for the time and $s,s'$ for a space direction. $a_\mu$ is the lattice 
spacing in direction $\mu$ (we write $\a_\mu$ if we want to consider the 
spacing as a vector in the positive $\mu$ direction), and we assume spatial 
isotropy, i.e.~that all spatial $a_\mu$ are the same, equal to $a_s$. For 
the extensions of the lattice we write $L_\mu$ and $N_\mu=L_\mu/a_\mu$ in 
physical and lattice units, respectively.}
i.e.~the product of link fields around a single plaquette (with $x$ chosen 
according to any consistent convention as one of its corners).
Expanding the action in terms of the field strength for small lattice spacings,
it is easy to see that at the classical level $\xi_0 = a_s/a_t$, if we use
$\beta\equiv 2N/g^2$ % for SU($N$) 
to introduce the bare coupling $g$      as in the isotropic case.
This justifies the name {\it bare anisotropy} for $\xi_0$.

Our problem is to determine the relation $\xi=\xi(\xi_0,\beta)$, or its inverse,
$\xi_0=\xi_0(\xi,\beta)$, which is actually preferrable from a 
practical point of view. In sect.~2 we describe the   % general
physical  ideas behind
the ``ratio method'' we use to determine this relation non-perturbatively.
Basically the same 
method was used previously in~\cite{QCDTARO,Scheideler,ScheidelerPhD}
(but not always with the same results, as we will see).
In sect.~3 we present
the details of our implementation of these ideas and of our simulations of
the anisotropic SU(3) plaquette action. Our results
cover      both fine and coarse lattices and
% TK2: four anisotropies in the range  $1.5 \leq \xi \leq 4$.
five anisotropies in the range  $1.5 \leq \xi \leq 6$.
In sect.~4 we use known perturbative results to motivate paramaterizations that
reproduce all our data within errors. We also
briefly discuss the application of our results in high precision thermodynamic
studies. We conclude in sect.~5 with a summary and an outlook on future work.

\section{Non-Perturbative Determination of the Renormalized 
         Anisotropy}\label{sec:nonpert}

The obvious spectral quantity to extract the renormalized anisotropy $\xi$ of
a gauge action from is the static potential. Actually, on an anisotropic 
lattice there are {\it two} potentials, according to whether the heavy quark 
and anti-quark propagate along the time or a space direction. We will refer to
them as the ``regular'' and ``sideways'' potential, and denote them by
$V_t(\r)$ and $V_s(\r)$, respectively.

On the lattice $V_t(\r)$ is measured in units of the temporal lattice
spacing and $V_s(\r)$ in terms of the spatial one. They therefore differ
by a factor of $\xi$. However, they also differ by an additive constant,
since the self-energy corrections to the static potential are different if 
the quarks propagate along the time or a space direction. To determine $\xi$
{}from a comparison of the regular and sideways potential we therefore have 
to demand,
\beq\label{xifromVst}
  V_s(\r) ~\stackrel{!}{=} ~ \xi \, V_t(\r) \+ \mbox{const} \, ,
 \qquad \mbox{for sufficiently large spatial}~\r \, .
\eeq
Note that due to the $\Ord(a_s^2,a_t^2)$ errors of the potential(s), the above
equality will in principle only hold for asymptotically large $r$.
{}From experience with the static potential we expect that the components of
$\r$ have to be at least $2-3$ in units of $a_s$, for the systematic
$\Ord(a_s^2,a_t^2)$ errors in $\xi$ to be below the 1\% level.

Due to the constant in~\eqn{xifromVst} this approach always requires several
$\r$ values to obtain an estimate of $\xi$. Since one cannot use small $r$ and
errors increase rapidly at larger $r$, this approach is relatively expensive
(and not very elegant, with a complicated error analysis if done properly).

A much better approach  involves comparing the sideways potential with the
quarks separated along a spatial direction with the case where they are
separated along the time direction.\footnote{We could also allow the separation 
to have a component along (another) spatial direction.}  
There is no additive constant to complicate matters and we can demand
\beq\label{xifromVs}
  V_s(y \a_s) ~\stackrel{!}{=} ~ V_s(y \xi \a_t) \qquad\quad
 \mbox{for sufficiently large}~y \, .
\eeq
Using  one's favorite method of calculating the static potential one can
obtain an estimate of $\xi$ for each value of $y$ (on the right hand side
one will generically have to interpolate between two values of $y\xi$).
The estimated $\xi$  should rapidly reach a plateau as $y$ increases.

This method is perfectly fine, in principle and in practice. If $V_s$ is
strictly meant to be the (sideways) static potential with negligible
finite-volume effects, it does however have a limitation. Namely, it will
be difficult to reach small lattice spacings,\footnote{As briefly mentioned in
the introduction,
we would like to reach small lattice spacings and make contact with perturbation
theory, so that we can present an analytic formula for the relation 
between $\xi$ and $\xi_0$ from weak to strong coupling.}
 where the elimination of
finite-volume effects in the potential becomes very expensive.
However, the finite-volume potential is also a spectral quantity;
if we can arrange the physical extent of the lattice to be the
{\it same} in the spatial and the time directions, then the finite-volume
effects in  $V_s(y \a_s)$ and $V_s(y \xi \a_t)$ should also be the same!
(Below we will describe one way of getting around the problem of how to 
choose the size of the lattice to correspond to the same physical extent 
in all directions without knowing $\xi$ beforehand.)

So, finite-volume effects should not prevent as from measuring $\xi$ on fine
lattices.  The remaining question is if we have a signal for ``times''
large enough so that ratios of Wilson loops allow us to extract the asymptotic
value of the (finite-volume) potential.  Actually, % we do not have to.
this is not necessary.
Similarly to the logic that allowed us to work with the finite-volume 
potential we note the following. If the ratios of Wilson loops that 
asymptotically would reach   $V_s(y \a_s)$ and $V_s(y \xi \a_t)$
have the {\it same} excited-state (i.e.~finite-``time'') contributions, we 
can use these ratios at finite ``times'' to measure $\xi$.
In other words, we have to set up a situation where the heavy quark anti-quark
states that can be thought of as underlying the spatial and temporal
Wilson loops in question 
correspond to the {\it same physical state} from a continuum point of view,
i.e.~when ignoring the usual $\Ord(a_s^2,a_t^2)$ lattice artifacts.

Before proceeding, let us introduce some notation for ratios of Wilson loops,
namely,
\beq\label{Rdef}
  R_{ss}(x,y) ~\equiv~ { W_{ss}(x,y)\o W_{ss}(x\!+\!1,y) } \, , \qquad\quad
  R_{st}(x,t) ~\equiv~ { W_{st}(x,t)\o W_{st}(x\!+\!1,t) } 
\eeq
in terms of spatial, $W_{ss}$, and temporal, $W_{st}$, Wilson loops.
To avoid cumbersome notation we here measure $x,y$ in units of $a_s$
and $t$ in units of $a_t$.   Asymptotically, for large $x$, the ratios 
 $R_{ss}(x,y)$ and $R_{st}(x,t)$  approach
$\exp[-a_s V_s(y \a_s)]$ and $\exp[-a_s V_s(t \a_t)]$, respectively.

For   $R_{ss}$ and $R_{st}$ to have the same 
excited-state contribution    %  is that all 
the spatial and temporal links making up the     Wilson loops in~\eqn{Rdef}
have to be     smeared by the same amount in physical units.
A simple way of achieving this is with {\it no smearing} at all. This leads
to the ``ratio-of-Wilson-loop'' method for determining $\xi$, which has been
used previously~\cite{QCDTARO,Scheideler,ScheidelerPhD}.
% (although it has never been emphasized why it works, even though one
% is not working with spectral quantities).

We will describe our detailed implementation(s) of this method in the
next section, where we also provide empirial support for our statements
concerning the cancellation of finite ``time'' and volume effects in
 $R_{ss}(x,y)$ and $R_{st}(x,t)$.

\section{Simulation and Results}\label{sec:results}

Let us start by summarizing our discussion of the ratio method in the 
previous section. Although the ratios of Wilson loops,
$R_{ss}(x,y)$ and $R_{st}(x,t)$ of eq.~\eqn{Rdef}, are {\it not} spectral
quantities (for $x$ where they can be measured reasonably accurately in
practice), we expect excited-state corrections to cancel between
them if $t\s= \xi y$ and all gauge links are unsmeared. Similarly,
finite-volume corrections to $R_{ss}(x,y)$ and $R_{st}(x,t)$ are the same
if the temporal and spatial extents of the box are equal in physical
units, i.e. $N_t\s= \xi N_s$ in lattice units.
Of course, these statements are expected to hold only if $x,y$ and $t$ are 
not too small; otherwise there can be large $\Ord(a_s^2,a_t^2)$ lattice 
artifacts.

The conceptually cleanest way to implement the above ideas in a simulation
is to turn the problem around and not measure $\xi$ for fixed $\xi_0$,
but rather try to determine the  $\xi_0$ that corresponds to a given $\xi$.
The following method is simplest if $\xi$ is an integer, so the reader might
want to have this case in mind for the moment. 

\begin{itemize}
\item  For fixed $\beta$ and $\xi$ choose a lattice volume 
$N_s^3\times N_t$ with $N_t\s= \xi N_s$ and calculate
\beq
%  \delta(x,y) ~\equiv~ \delta(x,y;t) ~\equiv~ \delta(x,y|\xi) ~\equiv~
%   { R_{ss}(x,y) \o R_{st}(x,t\!=\!\xi y) } \- 1
%
 \delta(x,y) ~\equiv~ \delta(x,y|\xi) ~\equiv~ \delta(x,y;t\s= \xi y) \, ,
 \qquad \mbox{where}\quad
  \delta(x,y;t) ~\equiv~ { R_{ss}(x,y) \o R_{st}(x,t) } \- 1 \, ,
\eeq
for two or three trial values of $\xi_0$.

\item  Interpolate in $\xi_0$ to find the zero-crossing of $\delta(x,y)$
for fixed $x,y$.  This determines an estimate of the non-perturbative
$\xi_0(\xi,\beta)$ for the given $x,y$.

\item  Consider different $x,y$; look for a plateau of $\xi_0(\xi,\beta)$ 
as $x$ and $y$ increase.

\end{itemize}

By starting at weak coupling, where perturbation theory can be used, and 
working one's way towards large coupling it is relatively easy to estimate 
trial values of $\xi_0$ that bracket the correct  $\xi_0(\xi,\beta)$ 
quite closely. In almost all cases we only needed two trial values for $\xi_0$.
The cases where simulations were performed at more than two values
were used to check that the $\delta(x,y)$ are sufficiently linear as 
functions of $\xi_0$ over the regions we typically had to interpolate.

We are mainly interested in integer $\xi$ for which the above method is best
suited. However, there is no problem in using it also for ``small fractions''.
For example, one case we investigated somewhat is $\xi={3\o 2}$. For even
$N_s$ almost everything goes through as above; the only change is that for
$\delta(x,y\s= 3)$, % TK2:
for example,
one has to perform an interpolation of
$\delta(x,y\s= 3;t\s= 4)$ and $\delta(x,y\s= 3;t\s= 5)\,$ to 
$\, t\s= \xi\! \cdot\! y\s= 4.5$.

As remarked earlier, the above method is perhaps the conceptually cleanest
way to proceed, since we choose $\xi$ not $\xi_0$ from the start and can
therefore make sure that $L_s\s= L_t$ in physical units. It was our method of
choice for all but the coarsest lattices.  However, on the coarsest lattices,
where one needs at least $1000$ configurations to measure $\xi_0(\xi)$ with 
some accuracy, one would like to avoid having to do independent simulations 
at two or more trial values of $\xi_0$ (on fine lattices one only needs a 
few hundred configurations, so this is not such an important issue). 
This can be achieved with a slightly modified procedure:

\begin{itemize}

\item Given a good guess for the value of $\xi_0$ corresponding to the desired
$\xi$, calculate $R_{ss}(x,y)$ and $R_{st}(x,t)$ by simulation at $\xi_0$
on an $N_s^3\times N_t$ lattice with $N_t\s= \xi N_s$.

\item For given $x,y$ interpolate  $R_{st}(x,t)$ in $t$ to match
      $R_{ss}(x,y)$. Estimate $\xi$ as $\xi \s= t/y$.

\item Find the plateau of the estimated $\xi$ as $x$ and $y$ increase.

\end{itemize}

As long as our estimate of $\xi_0$ was not too bad, all our previous remarks
about the cancellation of finite volume effects and such should apply.
%  and only insignificant systematic errors are introduced. 
In this method we obtain
$\xi(\xi_0,\beta)$, which can be translated into an estimate of 
$\xi_0(\xi,\beta)$ at the desired nearby $\xi$.   We have explicitly
checked in a few cases that both methods give consistent results.

We performed simulations at a large set of couplings aiming at $\xi=2,3,4$ and
% TK2: added xi=6:
a smaller set for $\xi=1.5, 6$. We generated between a few hundred and up to
% 2000 TK2:
2700 (almost) independent configurations at weak, respectively, strong coupling.  
We used two sets of code. One employs Metropolis, the other 
Kennedy-Pendleton~\cite{KPHB}  heatbath updating; 
in both cases alternating with microcanonical over-relaxation steps~\cite{CM}.

% TK2: moved this one paragraph up:
We started at weak coupling where our method worked so well that we initially
did not use link integration~\cite{PPR} for Wilson loops. 
On coarse lattices, however, fluctuations are large and link integration
yields a large saving in CPU time. It was implemented  by replacing 
independent links in a Wilson loop by their average over 20 local updates 
at fixed values of the staples surrounding the given link.

\begin{figure}[tbp]
\vskip -8mm
\mbox{ \hspace{-3.5em}\ewxy{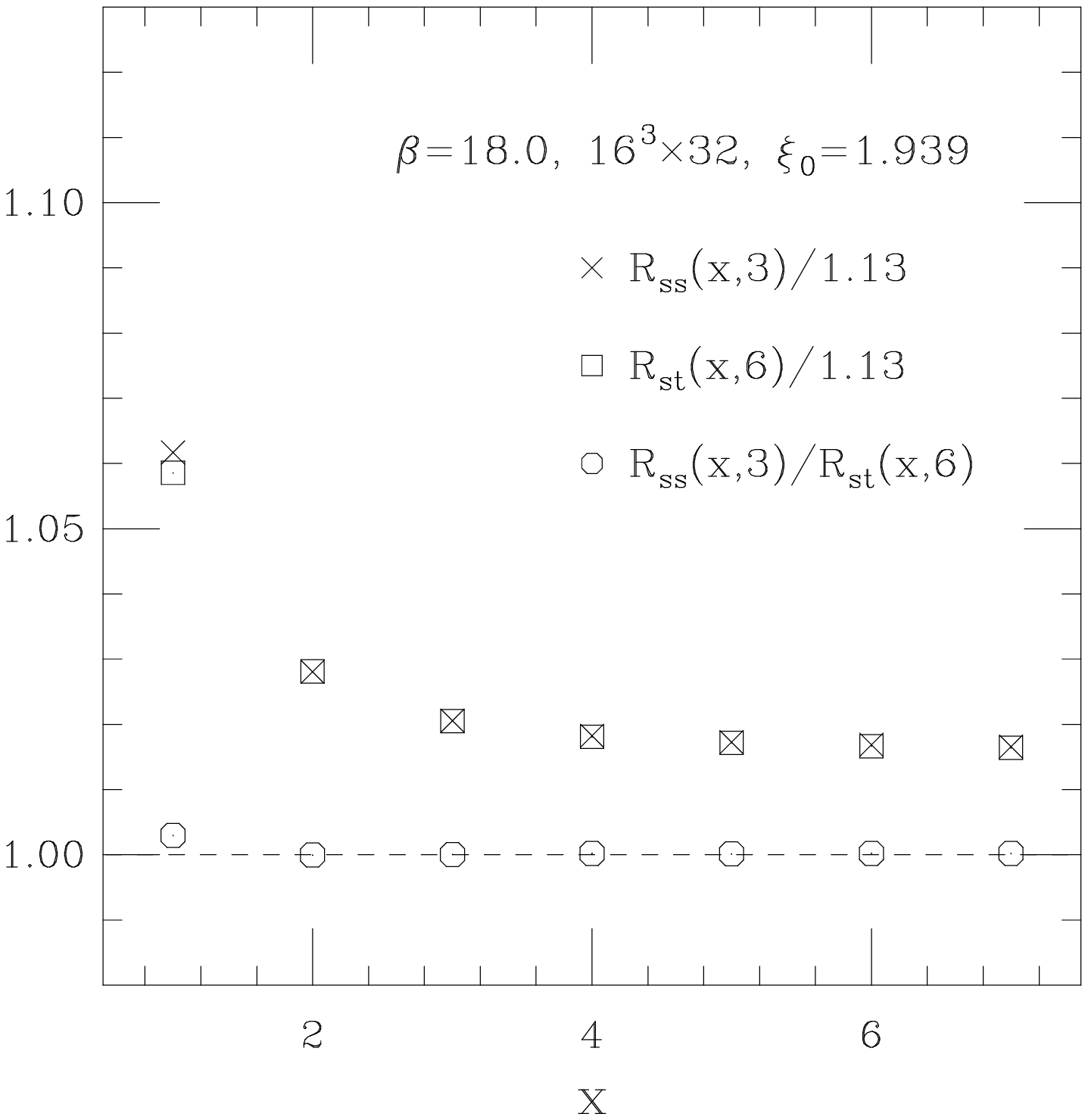}{100mm}   % was -3
       \hspace{-6.0em}\ewxy{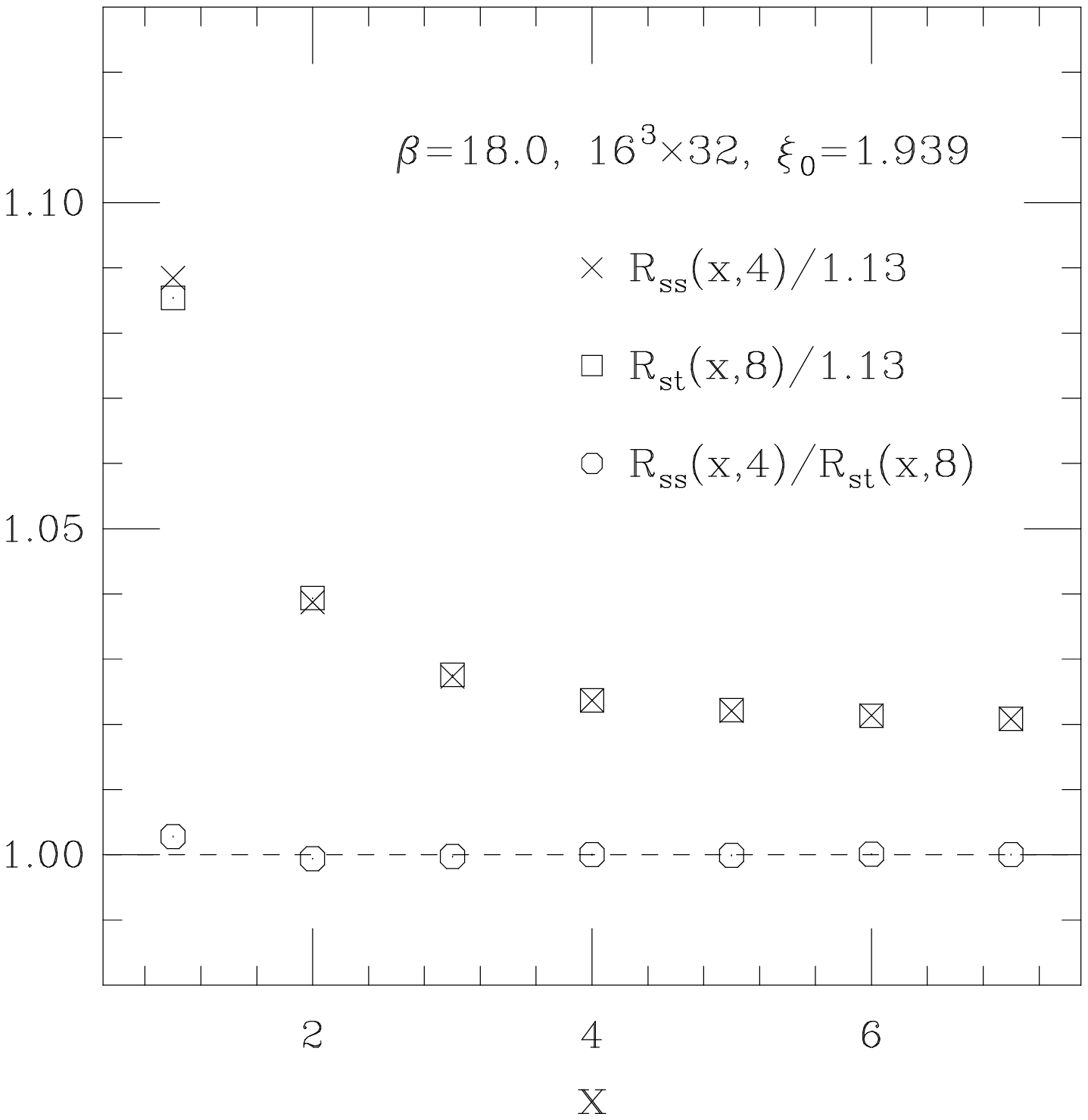}{100mm} } 
\vskip -10mm
\caption{Ratios $R_{ss}(x,y)$, $R_{st}(x,t)$ and the ``ratios of ratios''
         $R_{ss}(x,y)/R_{st}(x,t\s=\xi y)$ as functions of $x$ for fixed $y, t$
	 for a very fine lattice, $\beta=18$, $\xi\approx 2$.
         Shown are $y=3$ (left) and $y=4$ (right).
%	 Whereas the ratios of ratios reach their plateaux very early on both
%         very fine (left) and coarse (right) lattices, the individual ratios
%         $R_{ss}(x,y)$ and $R_{st}(x,t)$ are not close to their asymptotic
%         large-$x$ plateaux. 
         The $R_{ss}(x,y)$ and $R_{st}(x,t)$ have been
         rescaled by some overall factor simply so that they 
         can be put more meaningfully on the same plot with their ratio.
        }
\label{fig:plateau_R_RR_b18}
\vskip  3mm
\end{figure}

\begin{figure}[tbhp]
\vskip -8mm
\mbox{ \hspace{-3.5em}\ewxy{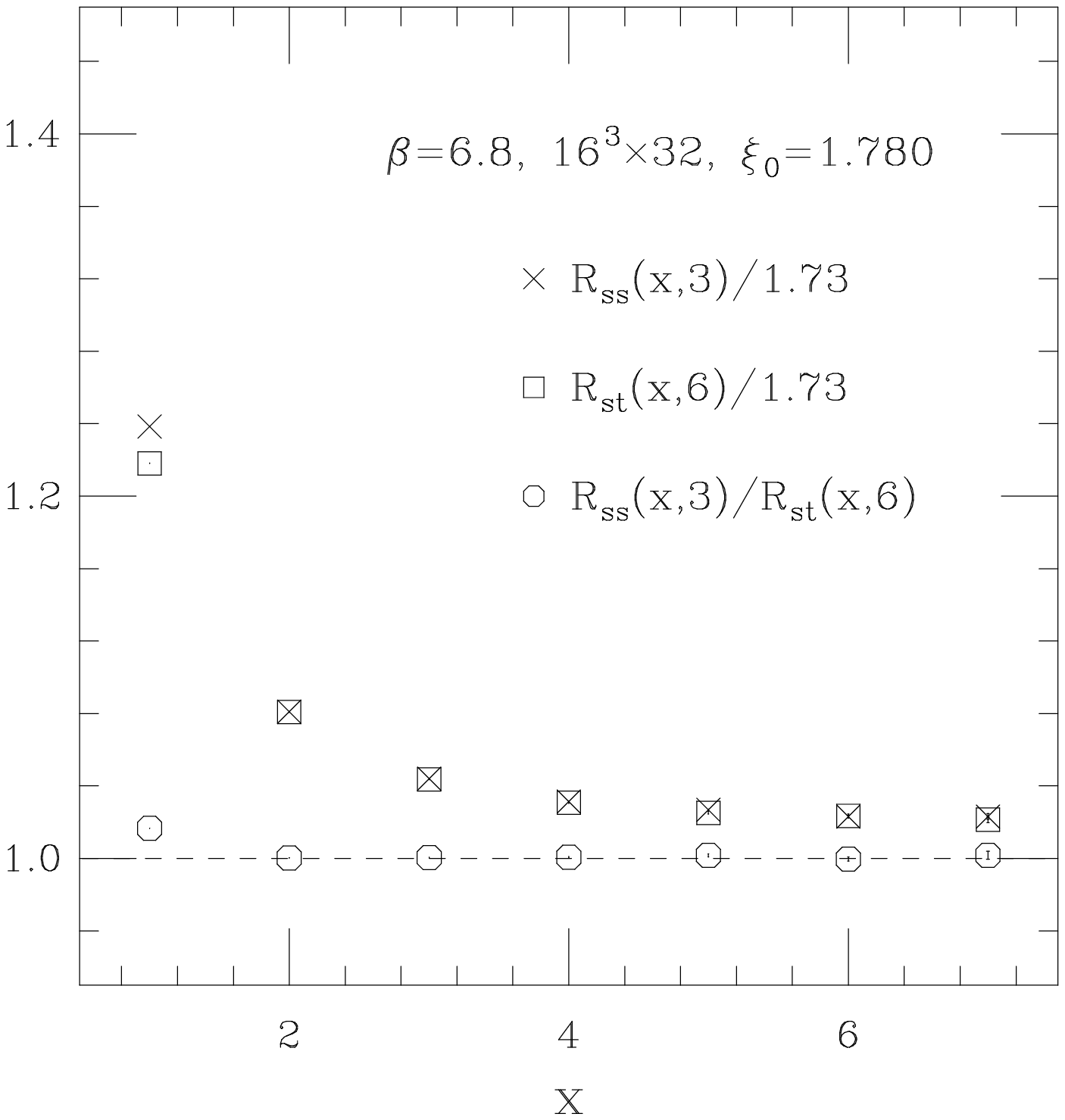}{100mm}   % was -3
       \hspace{-6.0em}\ewxy{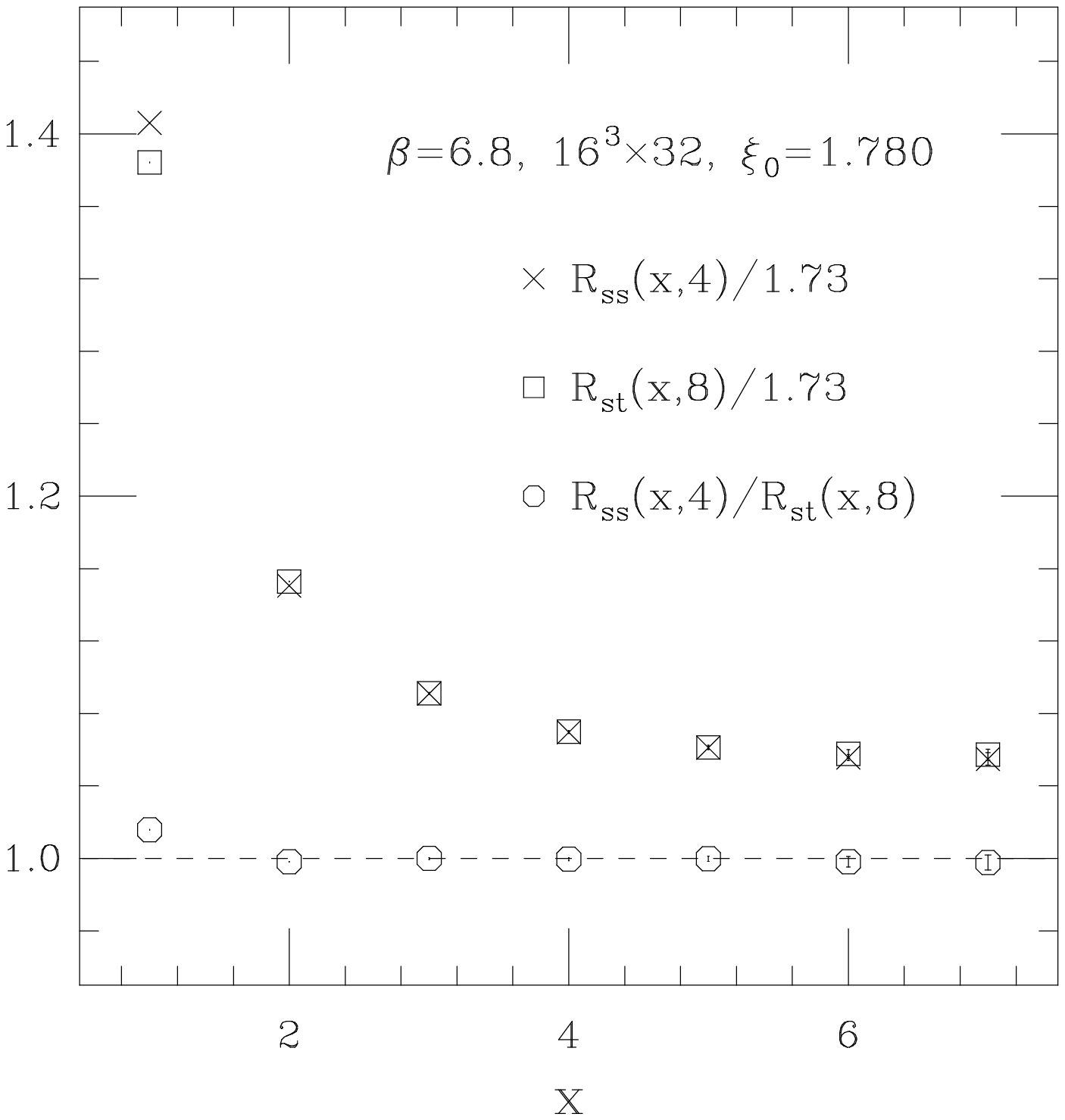}{100mm} } 
\vskip -10mm
\caption{As in figure~\protect\ref{fig:plateau_R_RR_b18} for an intermediate
	coupling, $\beta=6.8$, $\xi\approx 2$.
        }
\label{fig:plateau_R_RR_b68}
\vskip  3mm
\end{figure}

\begin{figure}[tbp]
\vskip -8mm
\mbox{ \hspace{-3.5em}\ewxy{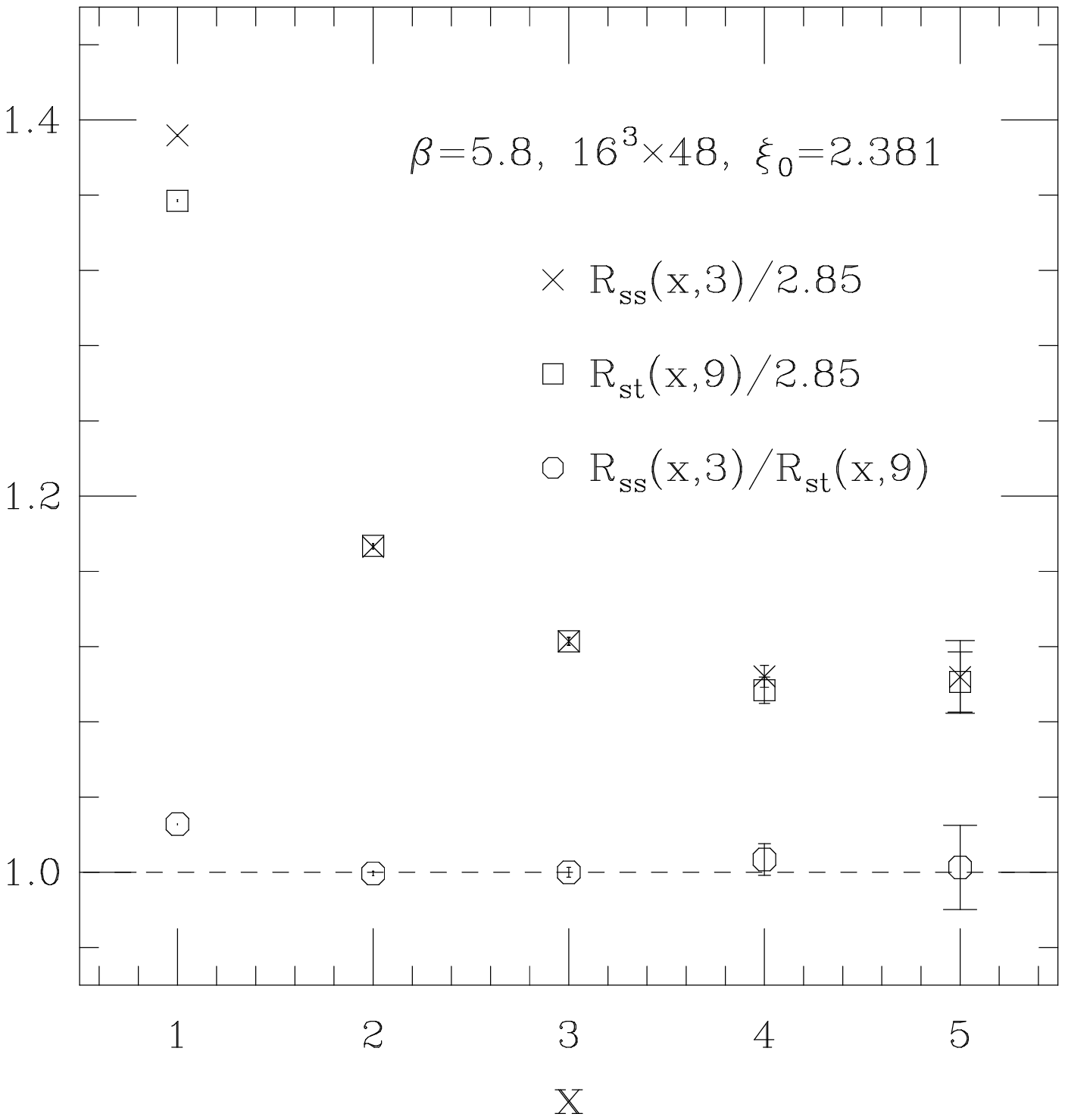}{100mm}   % was -3
       \hspace{-6.0em}\ewxy{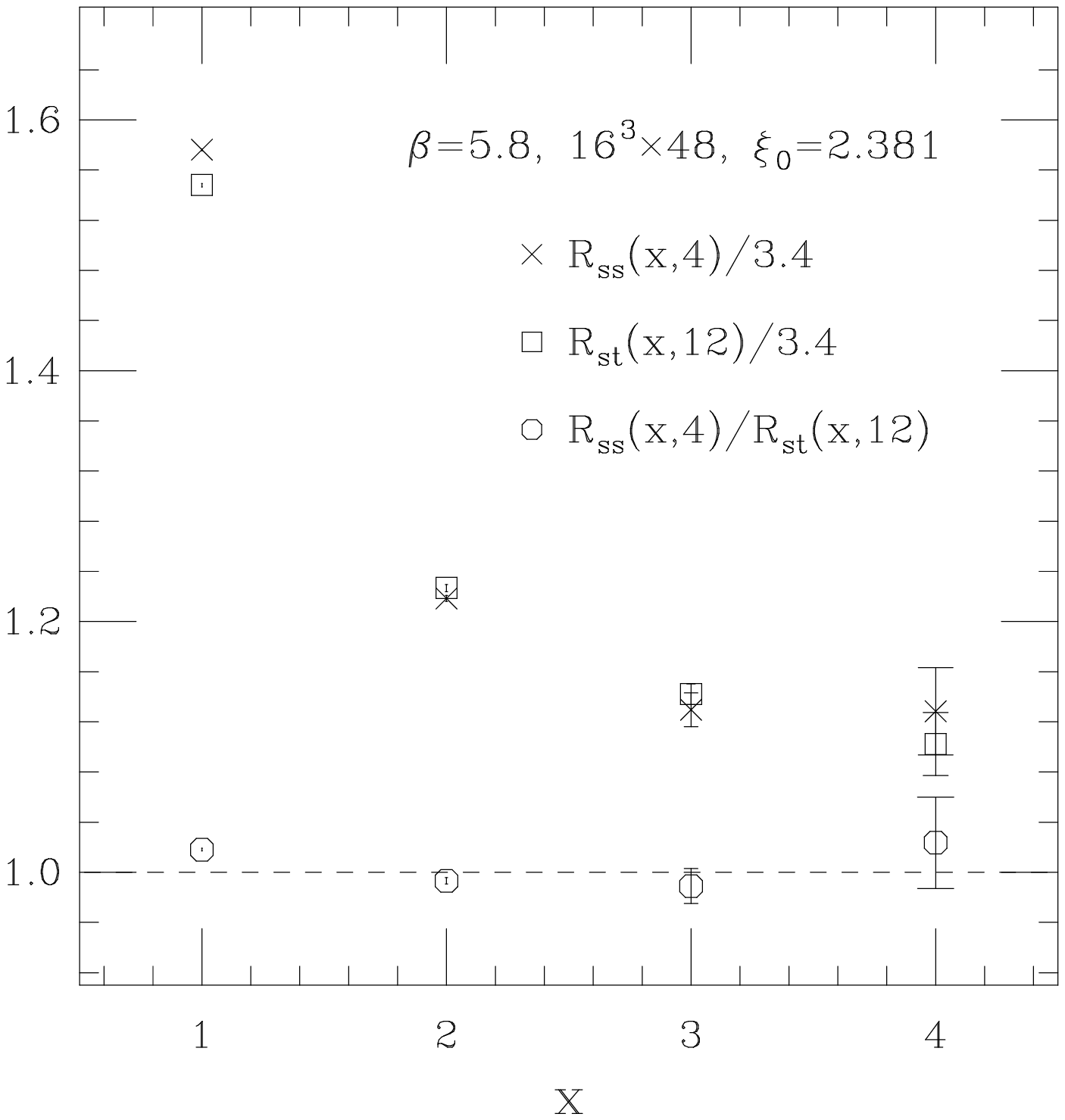}{100mm} } 
\vskip -10mm
\caption{As in figure~\protect\ref{fig:plateau_R_RR_b18} for a coarse lattice,
         $\beta=5.8$, $\xi\approx 3$.
        }
\label{fig:plateau_R_RR_b58}
\vskip  3mm
\end{figure}

\begin{figure}[tbph]
\vskip -8mm
\mbox{ \hspace{-3.5em}\ewxy{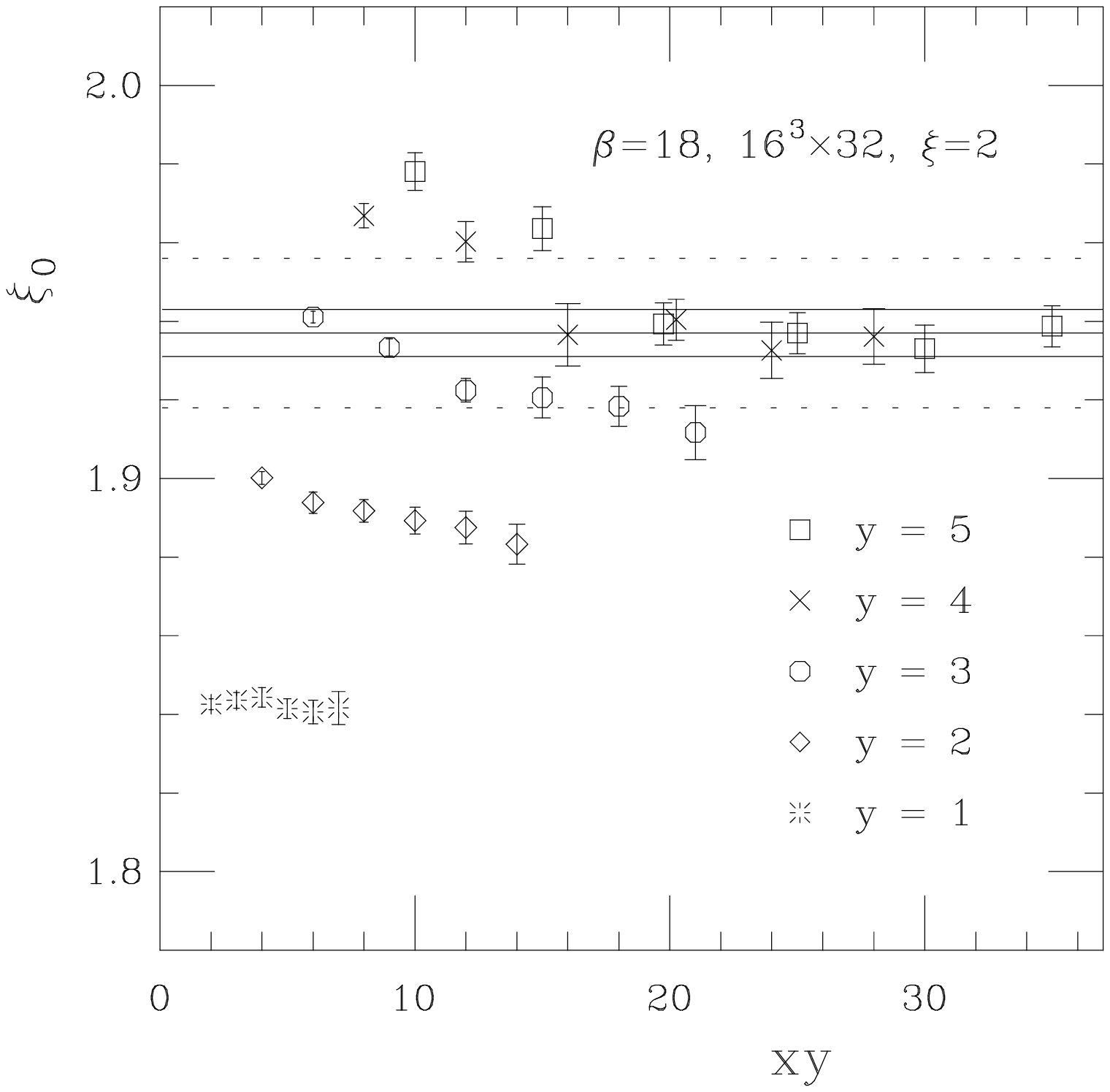}{100mm} 
       \hspace{-6.0em}\ewxy{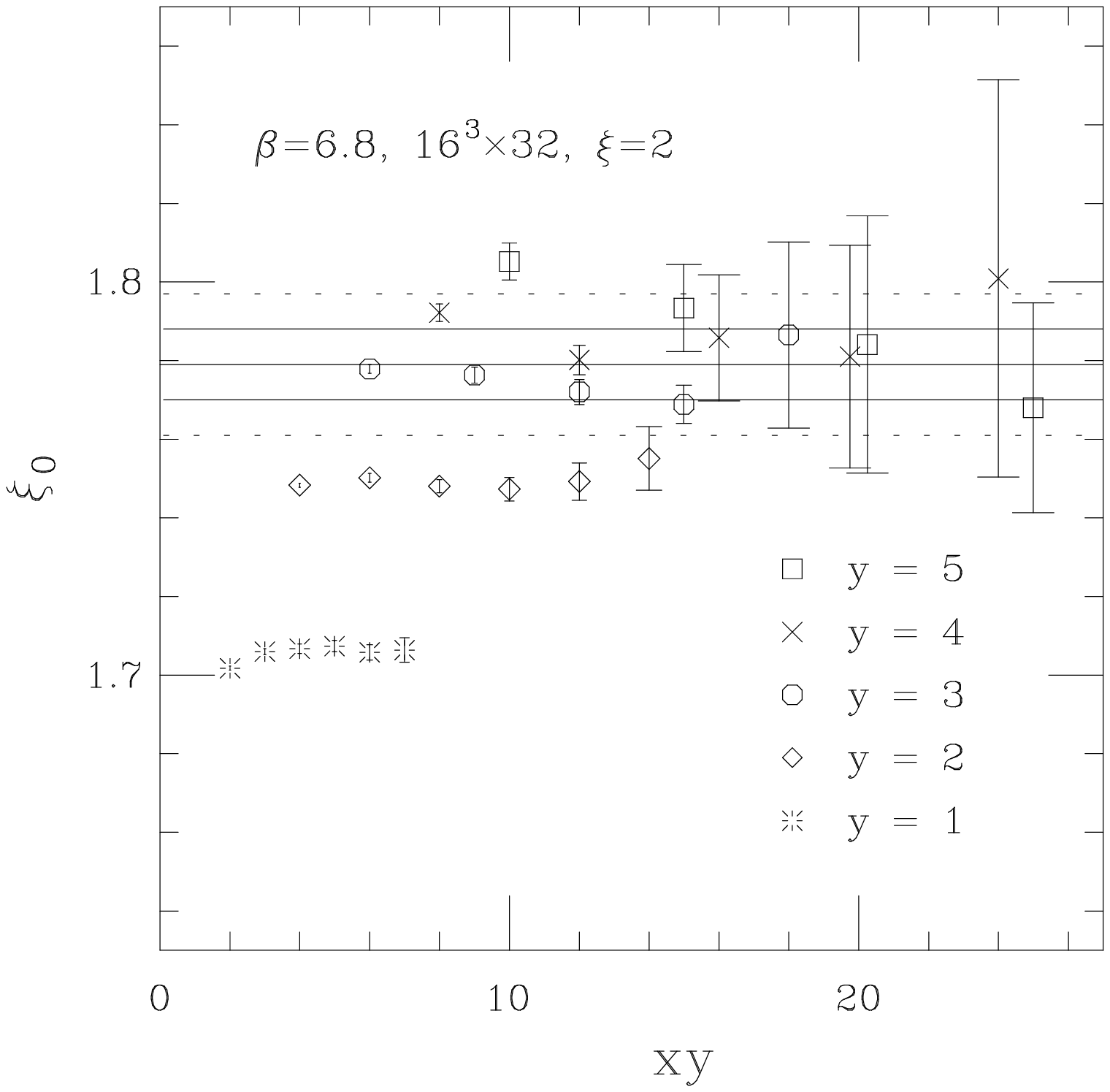}{100mm} } 
\vskip -10mm
\caption{$\xi_0$ estimated from different (ratios of) Wilson loops plotted
         versus the area of the Wilson loops. We show results for
         a very fine (left) and an intermediate (right) lattice.  
         Separate symbols
         are used to denote results from different $y$. The solid lines
	 are used to indicate our final estimate of $\xi_0$ and its error.
	 The short-dashed lines denote a 1\% band around the 
	 estimated $\xi_0$. Points that would otherwise overlap have been
	 slightly separated. For details see the main text.
        }
\label{fig:xi0_xy}
\vskip  3mm
\end{figure}

\begin{figure}[tbph]
\vskip -8mm
\centerline{\mbox{ \ewxy{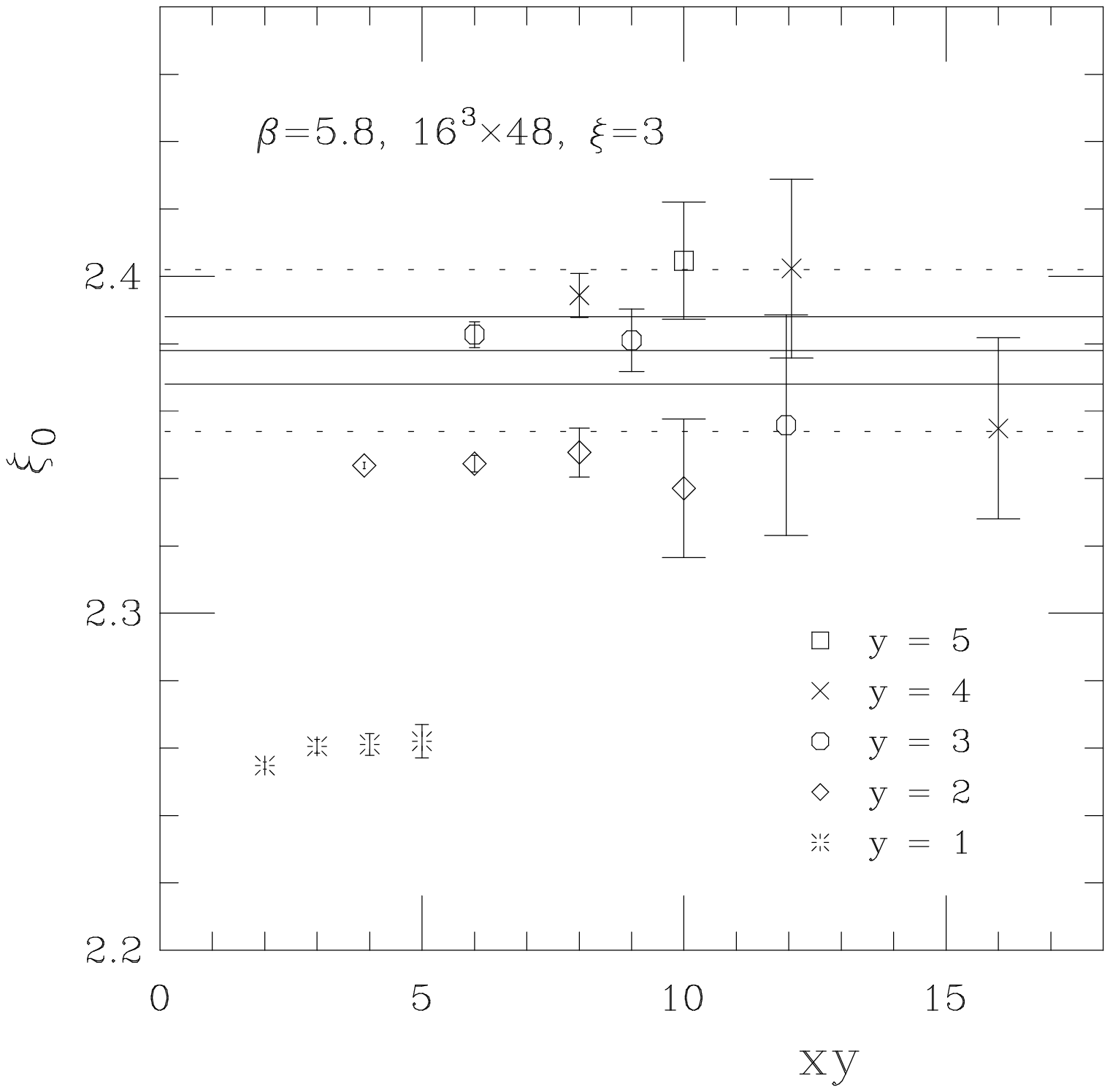}{100mm} } }
\vskip -10mm
\caption{As in figure~\protect\ref{fig:xi0_xy} for a coarse lattice,
	 $\beta=5.8$, $\xi= 3$.}
\label{fig:xi0_xy_2}
\vskip  3mm
\end{figure}

%  FINAL 31 data points into table:
%
%  dd := data_xi0;
%  for i from 1 to nops(dd) do
%    printf(`    & %4.1f   &    %6.4f(%.4f) \\\\ \n`, 
%            6/dd[i][1][1],dd[i][2],dd[i][3] );
%  od;

\begin{table}[tbph] \centering
\begin{tabular}{ | c | r | l | }
\hline
 ~~$\xi$~~ & ~~$\beta$~~ & ~~~$\eta$       \\    \hline
1.5 &  5.6   &   1.1462(114) \\  
    &  6.3   &   1.0981(48) \\  
    &  8.0   &   1.0631(43) \\  
    & 24.0   &   1.0183(35) \\  
2.0 &  5.4   &   1.2658(184) \\  
    &  5.5   &   1.2560(117) \\  
    &  5.6   &   1.2203(112) \\  
    &  5.8   &   1.1905(92) \\  
    &  6.3   &   1.1527(33) \\  
    &  6.8   &   1.1242(51) \\  % agrees with new 16^3 * 32 run !
    &  8.0   &   1.0941(30) \\  
    & 12.0   &   1.0515(44) \\ 
%%% TK2:  from 16^3 * 32:
    & 18.0   &   1.0325(32) \\  % NEW
    & 24.0   &   1.0214(26) \\  
3.0 &  5.5   &   1.3351(134) \\  
    &  5.6   &   1.3043(130) \\  
    &  5.8   &   1.2616(53) \\  % agrees with new 16^3 * 48 run !
    &  6.3   &   1.1947(38) \\  
    &  6.8   &   1.1650(36) \\  
    &  8.0   &   1.1186(33) \\  
    & 12.0   &   1.0676(30) \\  
    & 24.0   &   1.0288(18) \\  
4.0 &  5.4   &   1.4126(245) \\  
    &  5.5   &   1.3652(137) \\  
    &  5.6   &   1.3374(94) \\  
    &  5.8   &   1.2887(71) \\  
    &  6.3   &   1.2162(67) \\  
    &  6.8   &   1.1894(39) \\  
    &  8.0   &   1.1328(45) \\  
    &  9.5   &   1.1056(40) \\  
    & 12.0   &   1.0735(32) \\  
    & 24.0   &   1.0333(16) \\  
% TK2:
6.0 &  5.6   &   1.3738(114) \\
    &  6.3   &   1.2434(104) \\
\hline
\end{tabular}      
\vskip 2mm
\caption{Simulation results for the renormalization of the anisotropy,
          $\eta \s= \xi/\xi_0$.}
\label{tab:results}
\vskip 5mm
\end{table}

\vskip 1mm

Most simulations
were performed on lattices with $N_s\s= 8$ or $10$.  In several cases we also
ran at $N_s\s= 12, 16$, with otherwise the same parameters, and found
no significant difference from the results on smaller volumes. This is in accord 
with  the theoretical expectations discussed earlier. 

We performed one fine, one intermediate, and one coarse lattice simulation with 
$N_s\s= 16$, generating about the same number of configurations as for the
smaller volumes 
% TK2:
(we used three trial values of $\xi_0$ for the fine lattice, two for
the others).
 The errors of these results are therefore quite small;
they were used to investigate the question of when $\xi_0(\xi)$ determined
from different $\delta(x,y)$ reaches its plateau as $x$ and $y$ increase.
If we can show that the plateau is reached quite early,
% , i.e.~for relatively small $x,y$, 
we can then confidently determine it with less statistics in other cases.

First of all, we should point out the obvious, namely that the existence
of a ``plateau'' depends on the accuracy requested; if one demands
infinite precision, the plateau would of course only be reached as
$x, y\to \infty$.  Our aim is to determine the relation between
$\xi$ and $\xi_0$ to about 1\%. Let us now investigate when a plateau
of this accuracy is reached.

In figures~\ref{fig:plateau_R_RR_b18} $-$~\ref{fig:plateau_R_RR_b58}
we show the ratios $R_{ss}(x,y)$ and $R_{st}(x,t)$ and their
``ratio of ratios'' $R_{ss}(x,y)/R_{st}(x,t)$ as a function of $x$
for two different values of $y$. Clearly, excited-state contributions 
cancel in the ratio of ratios long before the individual ratios
reach their plateaux
in $x$  (in fact, it is not clear if they ever do for the unsmeared 
Wilson loops and volumes we use). This is in accord with expectations
discussed earlier. More quantitatively, it seems that $\delta(x,y)$ 
reaches its fixed-$y$ plateau to high precision when 
$x\geq {\rm max}(y\!-\!1,2)$. 
With somewhat less accuracy the plateau is apparently already reached
for $x\s= 2$, even for $y\geq 4$, at least for not too fine lattices.

Concerning the plateau of $\xi_0(\xi)$ % in $y$ 
consider figures~\ref{fig:xi0_xy} and~\ref{fig:xi0_xy_2}, 
where the values determined from different
Wilson loops are presented. For fixed $y$ we see the plateau emerging for
$x\geq {\rm max}(y\!-\!1,2)$. It seems that $y\s= 3$ is sufficient 
to reach the asymptotic plateau of $\xi_0$ within the accuracy of
interest. In fact, these and other results indicate that the
plateau can be obtained to 1\% or better from just
$\delta(2,3)$ and $\delta(3,3)$.\footnote{For fine lattices the $\xi_0$ 
determined from $\delta(x,3)$ for large $x>3$ 
might be     just on the boundary of the 1\% band around 
the asymptotic value we are aiming for. However, it would appear that
the $\xi_0$ from $x=2,3$ are well within the 1\% band, 
as figures~\ref{fig:xi0_xy} and~\ref{fig:xi0_xy_2}
and all our other results indicate.}
In all cases the value so obtained is consistent with that from
$y>3$ and $x\geq y-1$. Of course, at strong coupling the errors rise
so rapidly as $x,y$ increase, that it is % difficult to % rigorously 
next to impossible to
show with the same confidence and accuracy as on finer lattices that a plateau
has been reached. However, within the stated aim of 1\% errors our simulations
results indicate that this is the case. % However, one should realize that 
This is also plausible a priori, since
% $x,y=3$, say, correspond to {\it much} larger distances in physical units
% than on     finer lattices, where we could unambiguously see that such
% $x,y$ give results within 0.5\% or so of the asymptotic value.
$x,y=3$, say, which on fine lattices clearly gives results within 0.5\% or so 
of the asymptotic value, corresponds to a {\it much} larger distance in physical 
units on coarse lattices.

% to within $0.5\%$ or less (on coarse lattices
% the errors of determinations from $\delta(x,y)$ with larger $x,y$
% are of course much larger than this).

The final values we quote for $\xi_0(\xi)$ were obtained by not just
considering results from fixed $x,y$ (as in the figures above) 
but also {\it sums} of Wilson loops:
We sum $\delta(x,y)$ over various $x,y$ satisfying $y\geq 3$, $x\geq y\!-\!1$
and then find the zero crossing in $\xi_0$. Larger sums were considered for
fine lattices, smaller ones for coarse lattices. This was mainly done to
minimize the subjective element in choosing the final  $\xi_0(\xi)$. 
For the errors, on the other hand,  we were rather
conservative and take into account the scatter observed between 
different (sums of) Wilson loops. 

Errors of the individual $\delta(x,y)$ or sums thereof (and of $\xi=t/y$ 
in the second method)
were evaluated with the bootstrap method using
100 bootstrap ensembles. Before bootstrapping, the data were binned
to check for auto-correlations (and save disk space). We always found that
errors were stable, at least after some initial binning had been performed.
Our final results are shown in table~\ref{tab:results}, where we present
the measured {\it renormalization of the anisotropy}, $\eta\equiv \xi/\xi_0$.

Concerning the conceptual underpinnings of our method  we should
remark that we have investigated in one case what happens if one uses
{\it smeared} Wilson loops.  With no effort to tune the smearing to
be the the same in physical units (however one may want to define this)
when the (imaginary) heavy quarks 
are separated along a spatial or the temporal direction, we find that the
individual ratios $R_{ss}(x,y)$ and $R_{st}(x,t)$ reach their plateaux
earlier, as expected, but their ratio reaches its plateau {\it later} than
without smearing. This demonstrates that there is indeed a cancellation
of excited-state contributions in the ``ratio of ratios'' of unsmeared Wilson
loops.

\section{Parameterizing the Renormalization of the Anisotropy}\label{sec:param}

For future use in simulations we would like to present analytic
parameterizations of the results obtained in the previous section.
Given a good parameterization  it is guaranteed that
observables calculated  with the
anisotropic Wilson action extrapolate smoothly, like $a^2$
asymptotically, to the continuum limit. If we had performed simulations
at the central values of the measured $\xi_0(\xi,g^2)$ of the previous
section, data obtained from runs with sufficiently high statistics would
{\it not} lie on a smooth curve.  Furthermore, a good parameterization allows
us to perform simulations at any coupling, not just the ones we
happened to use in sect.~\ref{sec:results}.

\subsection{The $g^2$ dependence of $\eta$}

Obviously, a parameterization should be consistent with known
perturbative results.
The one-loop result for the renormalization of the anisotropy
$\eta$ of the Wilson gauge action was obtained long ago 
by Karsch~\cite{Karsch} (cf.~also~\cite{vBaal}).  At one loop $\eta$
can be written as, for SU($N$),
\beq\label{eta_oneloop}
 \eta ~=~ \eta(\xi_0,g^2) ~=~ 1 \+ {\eta_1(\xi_0)\o 2N}\, g^2 \+ \Ord(g^4) \, ,
\eeq
with $\eta_1(1)\s= 0$ and $\eta_1(\xi_0)$ increasing monotonically to some
finite value in the Hamiltonian limit $\xi_0 \to \infty$
(explicit numerical values for the $\eta_1(\xi_0)$ will be given below).
Note that to one-loop order we can replace $\xi_0$ by $\xi$ 
in this equation.  We can therefore incorporate the known
one-loop coefficients $\eta_1$ into a fit of the results of
sect.~\ref{sec:results} for fixed $\xi$.  

\begin{figure}[tb]
\vskip -6mm
\mbox{ \hspace{-3.5em}\ewxy{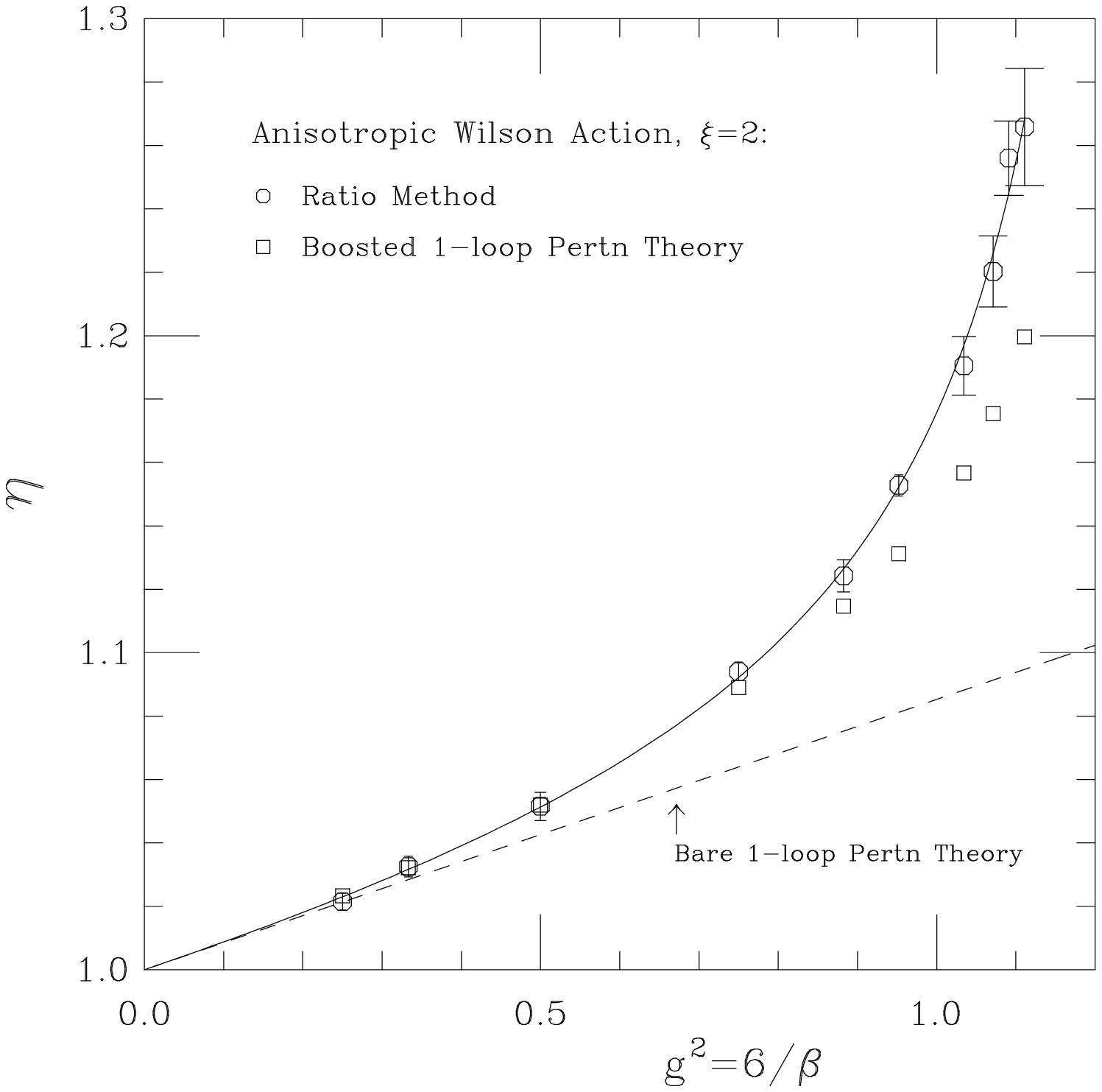}{100mm} 
       \hspace{-6.0em}\ewxy{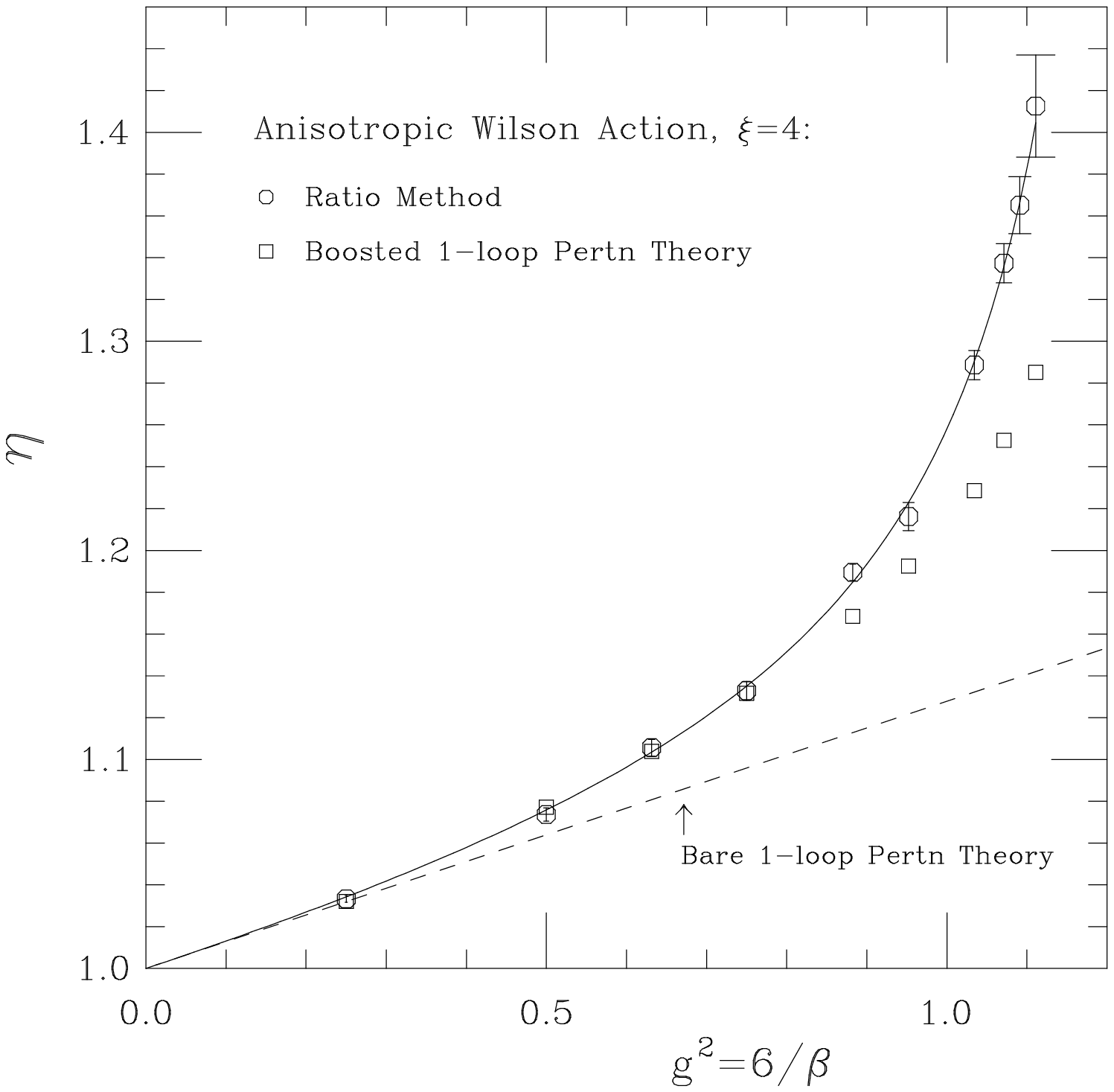}{100mm} } 
\vskip -10mm
\caption{Simulation results and fits of the renormalization of
         the anisotropy, $\eta$, 
         for fixed $\xi\s= 2$ (left) and $\xi\s= 4$ (right).
         For details see the main text.}
\label{fig:xi2and4}
\vskip  7mm
\end{figure}

We find that excellent fits of $\eta(g^2)$ for fixed $\xi$ are possible
to Pade ans\"atze of the form\
\beq\label{Pade}
  \eta(g^2) ~=~ {1+c_1\, g^2 + c_2\, g^4 \o 1 + c_0\, g^2 } \,\, ,
\eeq
with $c_1-c_0$ constrained to be consistent with one-loop perturbation theory,
i.e.~equal to $\eta_1(\xi)/6$.  The coefficients of parameterizations obtained
{}from fits to such  ans\"atze are given in table~\ref{tab:fits}, and a graphic
presentation of the $\xi=2$ and $4$ results is exhibited in 
figure~\ref{fig:xi2and4}. In these and later
plots ``boosted 1-loop perturbation theory'' 
refers to the replacement of the bare coupling $g^2$ by the ``boosted'' 
coupling~\cite{Par,LM}  
\beq
\tilde{g}^2 ~=~ { g^2 \o \sqrt{W_{ss}(1,1) W_{st}(1,1)} } \, .
\eeq
A more proper mean-field estimate would require~\cite{LM} an estimate of the 
appropriate scale for $\eta$, which presumably is much more infra-red than that
of the naive boosted coupling $\tilde{g}^2$. 
This would lead to better agreement with our 
non-perturbative determination at strong coupling.

\begin{table}[tb] \centering
\begin{tabular}{ | c | c | c | c | }
\hline
 $\xi$  &  $c_0$    &  $c_1$   &  $c_2$   \\ \hline
   2.0  &  -0.76216 & -0.67686 & -0.04351 \\
   3.0  &  -0.79245 & -0.67854 & -0.06625 \\
   4.0  &  -0.77715 & -0.64925 & -0.07038 \\
\hline
\end{tabular}      
\vskip 2mm
\caption{Coefficients of the parameterizations of 
         $\eta$ at fixed $\xi$, eq.~\protect\eqn{Pade}. }
\label{tab:fits}
\vskip 5mm
\end{table}

\vskip 1mm

\subsection{The $\xi$ dependence of $\eta$}

We are also interested in parameterizing our results for $\eta$ as a function 
of $\xi$ for fixed $\beta$, or, more generally, as a function of both
$\xi$ and $\beta$. To obtain an idea of how to proceed let us consider
what perturbation theory has to say about the $\xi$ dependence.
A plot of $\xi$ versus $\xi_0$ 
at one-loop using~\eqn{eta_oneloop} and the numerical values of
$\eta_1(\xi_0)$ from~\cite{Karsch} and~\cite{vBaal}
reveals that this relation is amazingly
linear, even at large coupling.  Rewriting~\eqn{eta_oneloop} as
\beq\label{eta_oneloopii}
 \xi ~=~ \eta \, \xi_0 ~=~ \xi_0 \+
  (\xi_0 - 1) \, {\hat{\eta}_1(\xi_0)\o 2N}\, g^2 \+ \Ord(g^4) \, , 
 \qquad\quad  
  \hat{\eta}_1(\xi_0) \equiv {\xi_0\o \xi_0 -1 } \, \eta_1(\xi_0) \, ,
\eeq
shows that this is equivalent to a very weak $\xi_0$ dependence of
$\hat{\eta}_1(\xi_0)$.
This is indeed the case, as can be seen in table~\ref{tab:etahat}.

\begin{table}[tb] \centering
\begin{tabular}{ | r | l | }
\hline
 $\xi_0$~~  & ~~$\hat{\eta}_1(\xi_0)$ \\ \hline
  1.00   & 1.00068(6) \\
  1.25   & 1.01116256 \\
  1.50   & 1.01758914 \\
  1.75   & 1.02143638 \\
  2.00   & 1.02364467 \\
  2.25   & 1.02480746 \\
  3.00   & 1.02516137 \\
  4.00   & 1.02319237 \\
  5.00   & 1.02090024 \\
  6.00   & 1.01886792 \\
  7.00   & 1.01716124 \\
  8.00   & 1.01574032 \\
 10.00   & 1.01354821 \\
 20.00  & 1.00846625 \\
$\infty$~ & 1.00250290 \\
%  above vBaal numbers have errors of less than 1 in last digit.
%
%  1.00     1.00068(6)
%  1.25     1.011162560
%  1.50     1.017589140
%  1.75     1.021436379
%  2.00     1.023644674
%  2.25     1.024807464
%  3.00     1.025161368
%  4.00     1.023192367
%  5.00     1.020900241
%  6.00     1.018867920
%  7.00     1.017161241
%  8.00     1.015740323
% 10.00     1.013548206
% 20.00     1.008466246
% $\infty$   1.002502899
\hline
\end{tabular}      
\vskip 2mm
\caption{One-loop coefficient of $\eta$ for the Wilson gauge
         action (see eqs.~\protect\eqn{eta_oneloopii} and~\eqn{eta_oneloop}
	 for the precise definition of $\hat{\eta}_1$).
         The $\xi_0\s= 1$ result is from~\protect\cite{Karsch}. All other
         values are from~\protect\cite{vBaal}; they should be accurate to
         the precision shown.}
\label{tab:etahat}
\vskip 5mm
\end{table}

The main $\xi_0$ dependence of the one-loop result is taken care of
by rewriting $\eta_1(\xi_0)$ in terms of $\hat{\eta}_1(\xi_0)$. We
have tried to parameterize the remainder by fitting $\hat{\eta}_1(\xi_0)$
to a Pade ansatz. 
We find that the results in table~\ref{tab:etahat} can be represented 
by
\beq
 \hat{\eta}_1(\xi_0) ~=~ { 1.002503\+ 0.39100\, y\+ 1.47130 {y}^{2} \- 
                           0.19231\, {y}^{3} \o 
       1\+ 0.26287\, y\+ 1.59008\, {y}^{2}\- 0.18224\, {y}^{3} }\, ,
 \qquad y \equiv {1\o \xi_0 } \, .
\eeq
This curve reproduces the numbers in table~\ref{tab:etahat} with an accuracy
of about  $10^{-6}$, which should be sufficient for all practical purposes.

We can now try to % TK2: fit  
find a representation of
all data for $\eta$. The most naive hope would be
that an ansatz of the form 
\beq\label{eta_all}
  \eta(\xi,g^2) ~=~ 1 \+ \left(1-{1\o \xi}\right)\, {\hat{\eta}_1(\xi)\o 6}\,
            \,    { 1 \+ a_1 \, g^2 \o 1 \+ a_0 \, g^2 } \,\, g^2 \, \, ,
\eeq
might be adequate. We find that this ansatz is not only adequate but
provides a nearly perfect representation 
of all data in table~\ref{tab:results} for $a_0=-0.77810$ and $a_1=-0.55055$
% TK2:
(the confidence level $Q$ of a fit is above 0.99, suggesting that our error
estimates  in sect.~\ref{sec:results} were indeed rather conservative).   
% The fit of these data has $\chi^2\s= 13.3$ with 29 degrees of freedom, 
% which implies a confidence level of $Q=0.99$ (this indicates that our error
% estimates  in sect.~\ref{sec:results} were indeed rather 
% conservative).     
It is quite remarkable that all our simulation
results can be summarized with just two fit parameters.
% For old:
% GAMMA(30/2,13.2894/2)/GAMMA(30/2) = 0.996392 with c0 frozen to opt value.
% So, true Q is:
% GAMMA(29/2,13.2894/2)/GAMMA(29.0/2) = 0.994424

\begin{figure}[tbp]
\vskip -8mm
\centerline{\ewxy{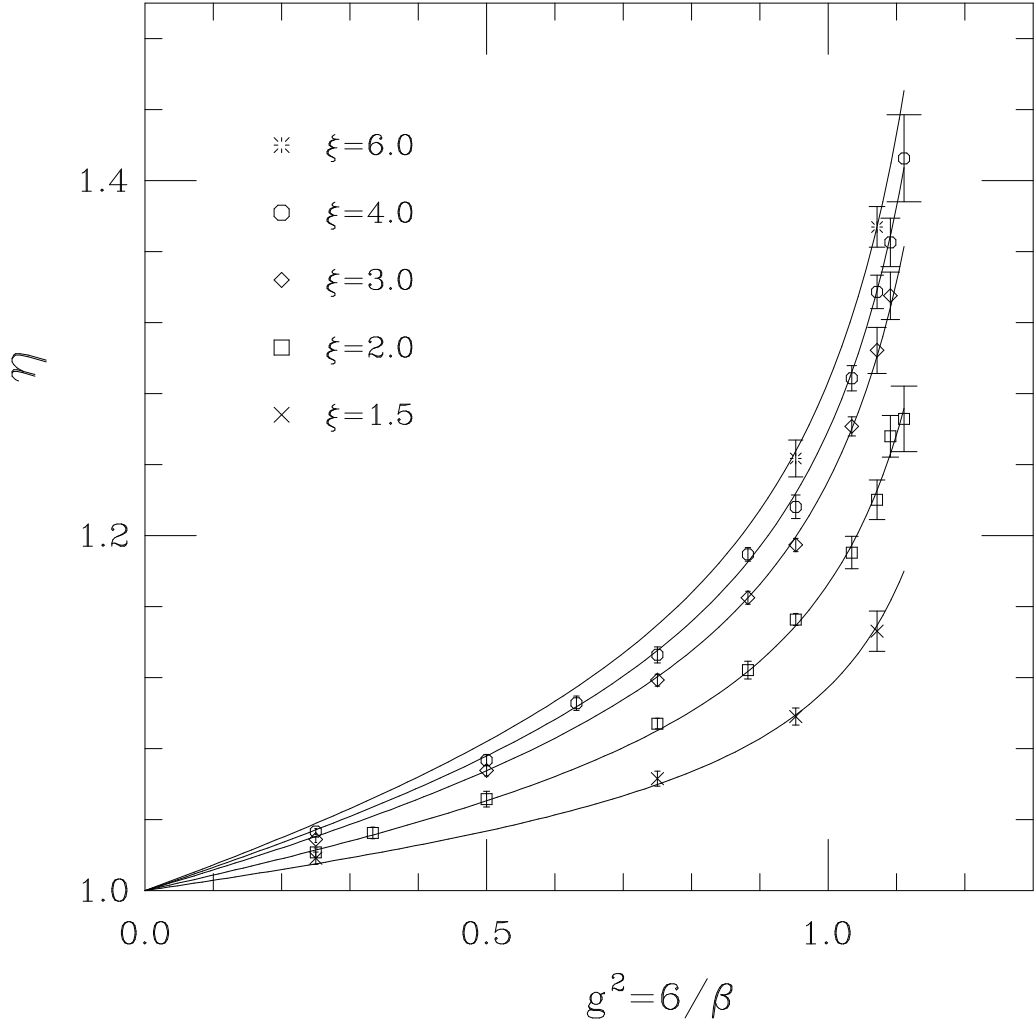}{200mm} }
\vskip -12mm
\caption{Fixed-$\xi$ cross-sections of our simulation results and the global 
         fit~\protect\eqn{eta_all} for the renormalization of the anisotropy,
         $\eta\s= \xi/\xi_0$. We show our fit up to $\beta\s= 5.4$.}
\label{fig:eta_g2}
\vskip 4mm
\end{figure}

\begin{figure}[tbp]
\vskip -8mm
\centerline{\ewxy{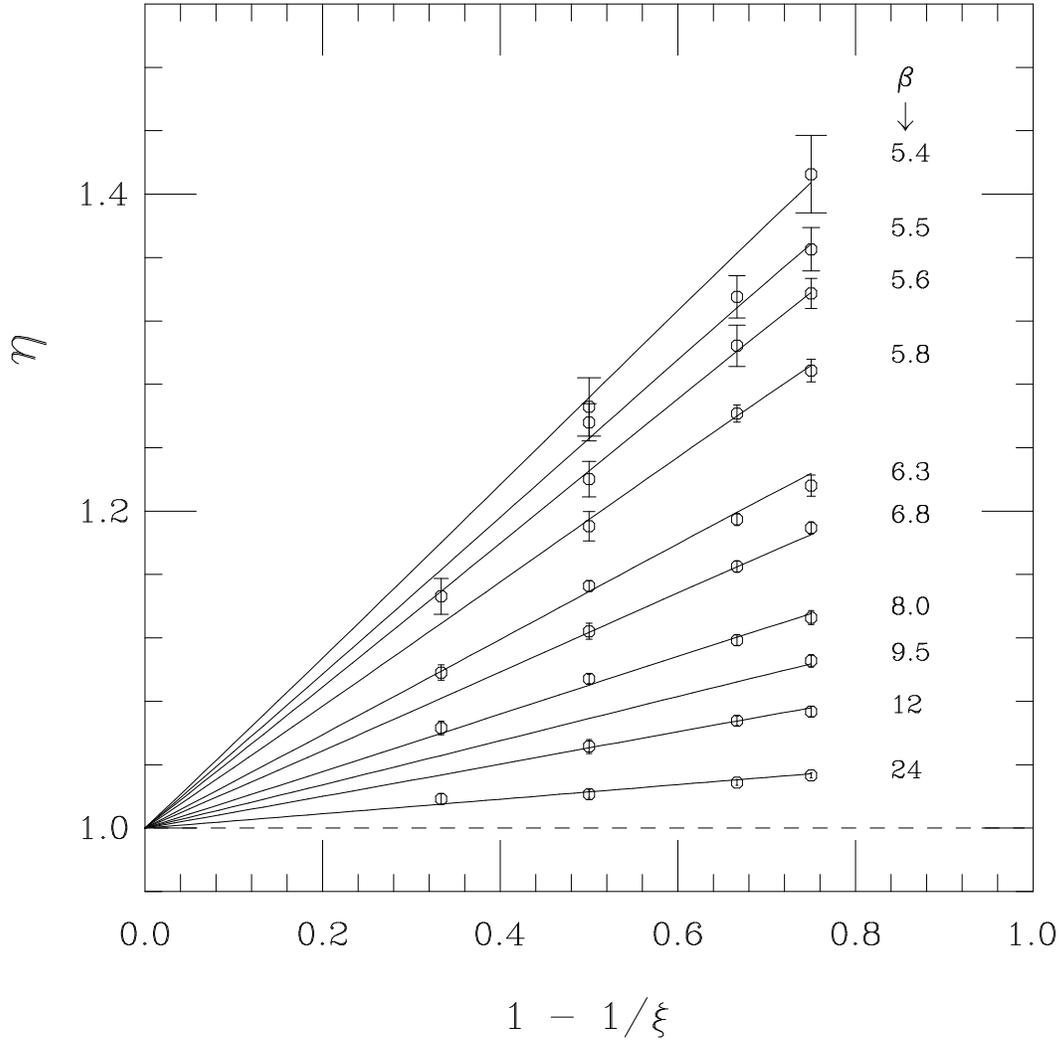}{200mm} }
\vskip -11mm
\caption{As in figure~\protect\ref{fig:eta_g2}, but now for fixed-$\beta$ cross 
	 sections.  For fixed $\beta$ the renormalization of the anisotropy
         $\eta$ appears to be an essentially linear function of $1-1/\xi$, 
         even on coarse lattices.}
\label{fig:eta_invxi}
\vskip 4mm
\end{figure}

Cross sections of the function~\eqn{eta_all} for
fixed $\xi$ and fixed $\beta$ can be found in figures~\ref{fig:eta_g2} 
and~\ref{fig:eta_invxi},  respectively.
Since there seems to be no systematic trend in the tiny deviations of the
curves
{}from the data, we recommend the use of~\eqn{eta_all} for future
% applications of the anisotropic Wilson action with $1 \leq \xi \leq 4$
%  TK2: 
applications of the anisotropic Wilson action with $1 \leq \xi \leq 6$
(instead of the fixed-$\xi$ representations  discussed in the previous
subsection).

\begin{figure}[tbph]
\vskip -8mm
\centerline{\ewxy{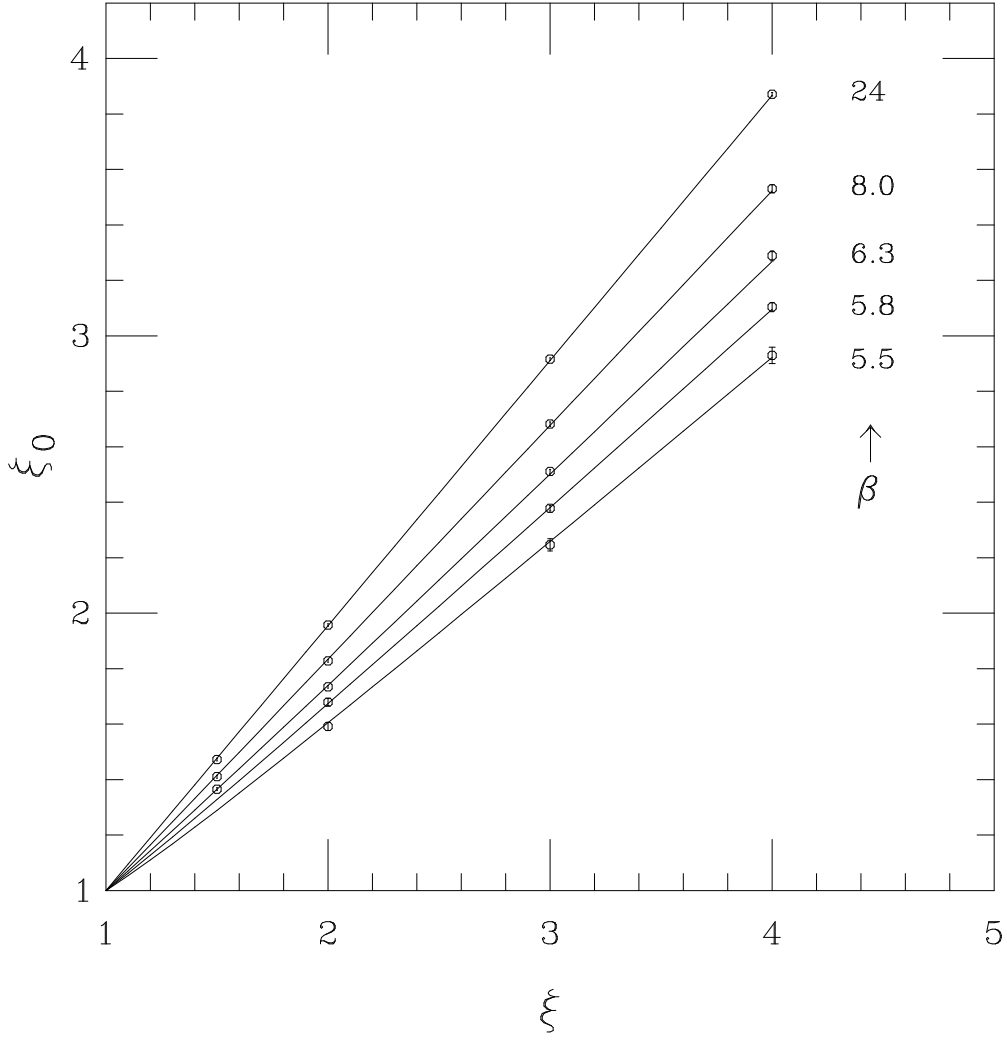}{200mm} }
\vskip -11mm
\caption{Similar to figure~\protect\ref{fig:eta_invxi}, but now we plot the bare 
	 versus the renormalized anisotropy (for clarity we only show a subset
         of the $\beta$ values we simulated). In this case deviations from 
         linearity are significant within our errors. For details see the main 
         text.}
\label{fig:xi0_xi}
\vskip 4mm
\end{figure}

%12345678901234567890123456789012345678901234567890123456789012345678901234567890
There is one interesting point concerning the $\xi$ dependence of 
eq.~\eqn{eta_all} that we should mention. Recall that in one-loop 
perturbation theory $\xi$ is a linear function of $\xi_0$ to high accuracy.
The same is therefore true for $\xi_0$ as a function of $\xi$. One might think 
that the accuracy of the ansatz~\eqn{eta_all} means that the same is true 
beyond the one-loop level. However, this is not the case. Trying to fit
$\xi_0$ as a linear function of $\xi$ for fixed $\beta$ usually leads to 
very small confidence levels (as low as 0.0005) for $\beta \leq 8.0$, whereas
quadratic fits always work very well.\footnote{To insure that the ansatz
for  $\eta$ is well-behaved
in the Hamiltonian limit, one might prefer to mean
$\xi_0 = \xi+ b_1(\xi-1)+ b_2(\xi-1)^2/\xi$  by ``quadratic ansatz'', 
instead of the naive $\xi_0 = \xi+ b_1(\xi-1)+ b_2(\xi-1)^2$.
However, this is not important here; either ansatz works fine for our
limited range of $\xi$.}
The non-linearity can be seen in figure~\ref{fig:xi0_xi} --- at least
with the help of a ruler ---  where we plot $\xi_0$ versus $\xi$ for
several $\beta$.
 This might sound paradoxical at first,
but the reason is that $\xi$ appears on the right hand side 
of~\eqn{eta_all}, not $\xi_0$, which might be considered the natural variable 
in perturbation theory (though it does not matter on the one-loop level).

First of all, the $\xi$ or $\xi_0$ dependence of $\hat{\eta}_1$ is  {\it not}
the problem; it is much too weak. The ansatz~\eqn{eta_all} works just
as well if we replace $\hat{\eta}_1$ by 1.02, say.
So we can write~\eqn{eta_all} as $\,\eta(\xi) = 1 + b z\,$, where
$z\s=(\xi-1)/\xi$ to a good approximation, and $b=b(g)$. This implies
$\xi_0 = \xi/\eta = \xi (1-b z+b^2 z^2 +\ldots)$. Our above finding
concerning fits of $\xi_0(\xi)$ therefore  essentially means this: a
``quadratic term'' (cf.~the previous footnote) is present,  but
within errors it is
indistinguishable from the term $b^2 z^2$ that arises from expanding 
$1/\eta$.      It is therefore simpler and much better to fit the
$\xi$ dependence of $\eta$ as in~\eqn{eta_all} instead of fitting
$\xi_0(\xi)$ to a quadratic ansatz (equivalently, one can fit $\xi_0(\xi)$
to the ansatz $\xi_0 = \xi/\eta$ with $\eta$ as above).

To summarize, the non-perturbative $\xi$ dependence of $\eta$ is
within our errors simply given by replacing $\xi_0$ by $\xi$ in the 
one-loop formula.  Why this should
be, we do not know, but we are certainly happy to exploit this fact in
presenting the simple parameterization~\eqn{eta_all} of all our data.

%  Q's of naive linear fits of xi0 versus xi at fixed beta:
%  beta     Q
%  24       0.707
%  12       0.668
%  8.0      0.045       vs 0.86  for quadr
%  6.8      0.596  ??
%  6.3      0.000476    vs 0.55  for quadr
%  5.8      0.15        vs 0.49  for quadr      
%  5.6      0.17        vs 0.92  for quadr
%  5.5      0.016       vs 0.85  for quadr

\begin{figure}[tbph]
\vskip -8mm
\centerline{\ewxy{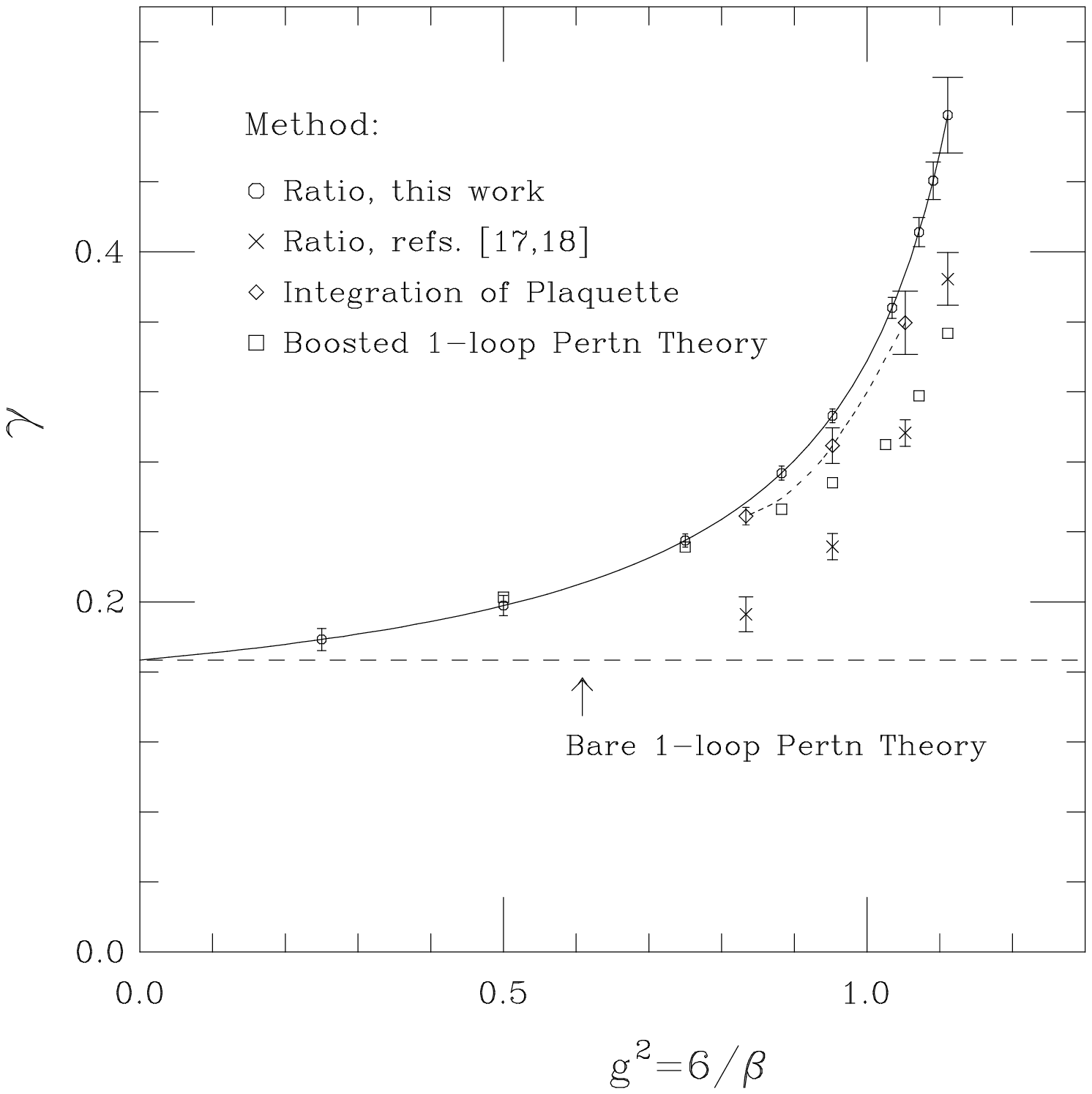}{200mm} }
\vskip -12mm
\caption{Comparison of non-perturbative determinations of
	$\gamma\equiv (1-\partial\xi_0/\partial\xi)/g^2$ at $\xi\s= 1$.
        The solid line denotes our result, eq.~\protect\eqn{gamma_g2},
        for $\beta \geq 5.4$, 
        the short-dashed line is the result from the integration of plaquette
        technique. The errors of the two methods are shown at several
        points. We also show results ($\times$) from a previous application of
       (a version of the) ratio method, which gave rather different results.}
\label{fig:dxi0_dxi}
\vskip 4mm
\end{figure}

\vskip 1mm
\subsection{The derivative $\partial\xi_0/\partial\xi$}

Among the immediate applications of this result we would like to point out
just one, related to thermodynamic studies. Obtaining all thermodynamic 
information requires one to take independent derivatives of the partition
function  with respect to temperature and volume. 
Even if one is ultimately interested the isotropic case, the most natural 
way to take these derivatives is to introduce
an anisotropic lattice with independent temporal and spatial lattice spacings
at an intermediate stage.  In addition to the
$\beta$-function for the dependence of the lattice spacing on the
coupling, one needs $\partial\xi_0/\partial \xi$ to calculate all
thermodynamic quantities (for details on lattice 
thermodynamics see e.g.~\cite{Engels,TDWil}). 

{}From eq.~\eqn{eta_all} one obtains
\beq\label{gamma_g2}
 {\partial \xi_0\o \partial \xi}(\xi\s= 1, g^2) ~=~ 
   1 \-  {\hat{\eta}_1(1)\o 6}\,
          { 1 \+ a_1 \, g^2 \o 1 \+ a_0 \, g^2 } \,\, g^2 
  ~\equiv ~ 1 \- \gamma(g) \, g^2 \, . 
\eeq
Our $\gamma(g)$ is shown in figure~\ref{fig:dxi0_dxi} together with results
{}from a different non-perturbative determination using the 
``integration of plaquettes'' technique~\cite{Engels,TDWil}
(we took the data from table~3 in~\cite{TDWil}).
This technique can not be used on very coarse lattices and becomes very
costly on fine lattices. We therefore only have results for 
$5.7\leq \beta\leq 7.2$ to compare with. No errors are quoted in~\cite{TDWil}
for these results, but we tried to estimate them from figure~1 
in~\cite{Scheideler} (the same information can be found with higher resolution
in figure~5.3 of ref.~\cite{ScheidelerPhD}, for example).
This error estimate is included in figure~\ref{fig:dxi0_dxi} for a few points.
For our results we also include errors, which
were obtained conservatively from fits of $\eta$ versus $\xi$ at
{\it fixed} $\beta$, instead of from our global fit~\eqn{eta_all}
(to be sure, the {\it central values} of our points in figure~\ref{fig:dxi0_dxi}
are from~\eqn{gamma_g2}).

Considering that different non-perturbative  methods  can differ
by $\Ord(a^2)$ lattice artifacts, the agreement between the two methods 
in figure~\ref{fig:dxi0_dxi} is excellent.
On the other hand, 
neither of these results agrees very well with estimates from a different 
application of the ratio method~\cite{Scheideler,ScheidelerPhD}, which is
also shown in figure~\ref{fig:dxi0_dxi}. For the largest $\beta$ 
considered these results also  appear to be in conflict with boosted 
perturbation theory.
(Our results here and other results~\cite{ALPHA,EHKlat97,EHKprl} have so
far always shown boosted perturbation theory to agree quite well with
non-perturbative determinations for $\beta\geq 7$ or so.)
The reason for this disagreement is not completely 
clear.\footnote{The most likely reason is this: In~\cite{Scheideler,ScheidelerPhD} 
the ratios $R_{ss}(x,y)$ and $R_{st}(x,\xi y)$
were not exactly matched, but rather were allowed to differ by an overall 
factor (the same for all ratios) that was then fitted and 
typically came out a few percent above 1 (unfortunately no errors were
quoted for this factor in~\cite{ScheidelerPhD}).
{}From our data we can extract the effect that the deviation of this 
factor (our ``ratio of ratios'') from its correct asymptotic value 1 has on the 
final $\eta(\xi,g^2)$.   It turns out that it
is sufficient to explain the quantitative differences between our and 
the results in~\cite{Scheideler,ScheidelerPhD}. In particular, for fine 
lattices $\eta(\xi,g^2)$ becomes rapidly more sensitive to deviations of the
ratio  of ratios from 1,
providing  an explanation for the fact that the difference
between our and the results of~\cite{Scheideler,ScheidelerPhD}
in figure~\ref{fig:dxi0_dxi} {\it increases} rather than decreases for small
coupling, contrary to naive expectations.}
% \footnote{In discussions with Tim Scheideler a possible reason emerged, however: 
% In~\cite{Scheideler,ScheidelerPhD} the ratios $R_{ss}(x,y)$ and $R_{st}(x,\xi y)$
% were not exactly matched, but rather were allowed to differ by an overall 
% factor (the same for all ratios) that was then fitted and 
% apparently did not always come out to be 1.}
% TK3: above after sending rev to NPB

% TK: added after sending to archive:
We should mention that in the SU(2) case the integration of plaquette
technique~\cite{EngelsSU2} and a different non-perturbative method~\cite{PGM}
also agree reasonably well (the latter giving slightly larger $\gamma(g)$), 
and are far above the perturbative result.
Qualitatively the situation is therefore the same as with our SU(3) results.

Using~\eqn{gamma_g2} and the $\beta$-function that can be obtained from the
accurate scale determinations for the isotropic
Wilson gauge action in~\cite{EHKsigma}
(and references therein), all ingredients required for high precision
thermodynamic studies with this action are now known for all lattice
spacings. Once the string tension  and/or Sommer scale $r_0$ has been determined
for the anisotropic Wilson actions, the same will hold for these actions.
Lattice artifacts in the thermodynamics of the anisotropic Wilson (gauge or quark) 
action are considerably smaller~\cite{anisoTD,Kacz} than for the isotropic case, 
so it might be interesting to pursue such studies on anisotropic lattices.

\vskip 2mm

\section{Conclusion and Outlook}
% \vskip 1mm

Using a method based on ratios of Wilson loops we have presented an accurate,
non-perturbative determination of the relation between the bare ($\xi_0$) and
renormalized ($\xi$) anisotropies of the Wilson gauge action. We have argued that
these ratios should be thought of as the ``finite-volume static potential with
excited-state contributions''. By creating the same physical situation with the
heavy quarks separated either along a spatial or a temporal direction there
should be no significant finite-volume or excited-state corrections to the 
relation between $\xi_0$ and $\xi$   even for moderately small Wilson loops.
 We have explicitly seen this in our simulations.
This fact makes our method quite cheap.
Nothing in our method is specific to the Wilson action, so it can also be
be applied to improved gauge actions.

Our results for $\xi_0(\xi,g^2)$ are significantly more accurate than previous
ones, were obtained at a fraction of the cost, and are the only ones that cover
the full range from weak to strong coupling\footnote{Preliminary results 
indicate that the lattice spacing
for $\beta\s= 5.5$ and $\xi\geq 2$ is close to 0.3~fm. The range of
couplings we covered should therefore be sufficient for all future
applications.}
and a large range of anisotropies.
Furthermore, we have presented a simple parameterization of our results,
eq.~\eqn{eta_all} with $a_0=-0.77810$ and $a_1=-0.55055$, that
reproduces all data within errors and is consistent with perturbation theory
at weak coupling. 
% TK:  for NPB:
For lattice spacings of interest in practice
we find that   the renormalization of the anisotropy 
               is much larger than predicted
by one-loop perturbation theory % TK2: (even with mean-field improvement).
(even when ``boosted'').
Our results are therefore crucial for future applications.

% TK2: 
The errors of our results,   about 1\% on coarse and less on finer
lattices, should also be  sufficient for future applications.
Since our parameterization incorporates the known one-loop 
behavior, it is % virtually guaranteed 
quite clear that the error in an observable
due to the remaining small error in $\xi_0(\xi,g^2)$ will (almost completely)
extrapolate away in the continuum limit.\footnote{We are here referring to the 
statistical and systematic errors of {\it our method} of determining 
$\xi_0(\xi,g^2)$;  another legitimate method can of course differ by O($a^2$) 
terms, which are a priori known to extrapolate away in the continuum limit.}

% Our results make 
Given these results, the anisotropic Wilson gauge action    is   
               as simple to use as the
isotropic one. The only thing that is currently missing are accurate 
determinations of the physical scale (string tension and Sommer scale $r_0$) 
of the anisotropic actions. 
This gap should soon be filled, since once the renormalized anisotropy is known, 
the ``regular'' static potential (cf.~sect.~\ref{sec:nonpert}) can be
determined very accurately. Work on this is in progress.

A number of projects using the anisotropic Wilson action can now  
start immediately. One example is the study of glueballs in pure gauge 
theory. It will be interesting to see how much of the improvement seen in
glueball studies with anisotropic {\it improved} actions~\cite{MorPea} is 
due to the anisotropy and how much due to the elimination of (most of) the
O($a_s^2$) errors. 
                    Another example, as
remarked at the end of sect.~4, would be to reconsider
high precision thermodynamic studies using our results.

One could also % initiate % studies % calculations  
start simulations  of heavy quark systems in the quenched approximation. 
An immediate improvement compared to isotropic 
studies should be apparent (cf.~\cite{AKLlat96}).
However, as mentioned in the introduction,
 truly reliable and accurate studies of such systems will have to 
await the non-perturbative          O($a^0$)  and
O($a$) improvement of Wilson-type
quark actions on anisotropic lattices: 
The first (and easier) step is to tune the {\it bare velocity of light} 
of the quarks so that the fermion and gauge sectors agree on the 
renormalized anisotropy.    The second step involves the tuning of
the temporal and spatial clover coefficients to eliminate O($a$) 
errors~\cite{SF}.\footnote{This presumably has some (small) effect on the
bare velocity of light, which therefore has to be retuned iteratively
with the clover coefficients. We expect this iterative retuning to
converge rapidly, if necessary at all.}

% As remarked at the end of sect.~4, it might also be interesting to reconsider
% high precision thermodynamic studies in light of our results.

Finally, since our method is free of most lattice artifacts
whose elimination would be very expensive beyond the quenched approximation,
the prospects for extending this work to full QCD look good. Of course,
simulations for full QCD will  
be significantly more expensive, not just because the determinant of the 
quark matrix has to be calculated, but also because one has to tune more bare
parameters simultaneously  to obtain consistent quark and gauge anisotropies
(now the quark parameters feed back into the gauge sector).
However, the required tuning    might not be as hard as it first sounds.

\vskip 10mm
% \clearpage

\noindent
{\bf Acknowledgements}
\vskip 2mm

\noindent    
I would like to thank Tim Scheideler for discussions and correspondence,
as well as Mark Alford and Urs Heller for comments on the manuscript.
This work is supported by DOE grants DE-FG05-85ER25000 and 
DE-FG05-96ER40979.
% The computations in this work were performed on the workstation
Most of the computations in this work were performed on the workstation
cluster at SCRI; the three simulations on $N_s\s=16$ lattices were performed
on the QCDSP supercomputer at SCRI.
 Two different sets of code were used in this
project. One was developed in collaboration with Mark Alford and
Peter Lepage, the other is SZIN, a macro-based C package 
% for QCD simulations
developed at SCRI for QCD simulations on a variety of platforms.

\newpage

%%%%%%%%%%%%%%%%%%%%%%%%%%%%%%%%%%%%%%%%%%%%%%%%%%%%%%%%%%%%%%%%%%%%%%%%%
% \appendix
% \section{Notation and Conventions}\label{app:notation}
% \subsection*{A.1~~~~In the Continuum}
% \section{}

\end{document}